\documentclass{aastex631}

\usepackage{placeins}
\usepackage{subfigure}

\usepackage{longtable}
\usepackage{booktabs}
\usepackage{natbib}
\usepackage{tabularx}
\usepackage{array} % For defining column types
\received{October 5 , 2024}
\revised{December 13, 2024}
\accepted{January 23, 2025}
\begin{document}

%\title{Template \aastex Article with Examples: 
%v6.3.1\footnote{Released on March, 1st, 2021}}

\title{The Ophiuchus DIsk Survey Employing ALMA (ODISEA): Complete Size Distributions for the 100 Brightest Disks Across Multiplicity and SED Classes}

\author[0009-0009-8115-8910]{Anuroop Dasgupta}
\affiliation{Instituto de Estudios Astrofisicos, Facultad de Ingeniería y Ciencias,
Universidad Diego Portales,  Av. Ejercito 441, Santiago, Chile}
\affiliation{Millennium Nucleus on Young Exoplanets and their Moons (YEMS), Chile}

\author[0000-0002-2828-1153]{Lucas A. Cieza}
\affiliation{Instituto de Estudios Astrofisicos, Facultad de Ingeniería y Ciencias,
Universidad Diego Portales,  Av. Ejercito 441, Santiago, Chile}
\affiliation{Millennium Nucleus on Young Exoplanets and their Moons (YEMS), Chile}

\author[0000-0003-4907-189X]{Camilo Gonz\'alez-Ruilova}
\affiliation{Departamento de Física, Universidad de Santiago de Chile, Av. Victor Jara 3659, Santiago, Chile}
\affiliation{Millennium Nucleus on Young Exoplanets and their Moons (YEMS) }
\affiliation{Center for Interdisciplinary Research in Astrophysics Space Exploration (CIRAS), Universidad de Santiago de Chile, Chile}

\author[0000-0002-4314-9070]{Trisha Bhowmik}
\affiliation{Instituto de Estudios Astrofisicos, Facultad de Ingeniería y Ciencias,
Universidad Diego Portales,  Av. Ejercito 441, Santiago, Chile}
\affiliation{Millennium Nucleus on Young Exoplanets and their Moons (YEMS), Chile}

\author[0000-0003-2406-0684]{Prachi Chavan}
\affiliation{Instituto de Estudios Astrofisicos, Facultad de Ingeniería y Ciencias,
Universidad Diego Portales,  Av. Ejercito 441, Santiago, Chile}
\affiliation{Millennium Nucleus on Young Exoplanets and their Moons (YEMS), Chile}
\author[0009-0007-4878-0252]{Grace Batalla-Falcon}
\affiliation{Instituto de Estudios Astrofisicos, Facultad de Ingeniería y Ciencias,
Universidad Diego Portales,  Av. Ejercito 441, Santiago, Chile}

\author[0000-0002-7154-6065]{Gregory Herczeg}
\affiliation{Kavli Institute for Astronomy and Astrophysics, Peking University, Beijing 100871, China}
\affiliation{Department of Astronomy, Peking University, Beijing 100871, China}
%\affiliation{Visiting astronomer, Department of Astronomy; California Institute of Technology; Pasadena, CA 91125, USA}

\author[0000-0003-3573-8163]{Dary Ruiz-Rodriguez}
\affiliation{National Radio Astronomy Observatory, 520 Edgemont Road, Charlottesville, VA 22903-2475, United States of America}

\author[0000-0001-5058-695X]{Jonathan P. Williams}
\affiliation{Institute for Astronomy, University of Hawaii, Honolulu, HI 96822, USA}

\author[0000-0002-5991-8073]{Anibal Sierra}
\affiliation{Mullard Space Science Laboratory, University College London, Holmbury St Mary, Dorking, Surrey RH5 6NT, UK}

\author[0000-0002-0433-9840]{Simon Casassus}
\affiliation{Departamento de Astronomía, Universidad de Chile, Casilla 36-D, Santiago, Chile }
%\affiliation{Data Observatory Foundation, to Data Observatory Foundation, Eliodoro Yañez ˜ 2990, Providencia, Santiago, Chile}
\affiliation{Millennium Nucleus on Young Exoplanets and their Moons (YEMS), Chile}

\author[0000-0001-8577-9532]{Octavio Guilera}
\affiliation{Instituto de Astrofísica de La Plata (IALP), CCT La Plata-CONICET-UNLP, Paseo del Bosque s/n, La Plata, Argentina}
%\affiliation{N´ucleo Milenio de Formaci´on Planetaria (NPF), Chile.}

\author[0000-0002-5772-8815]{Sebastian P\'erez}
\affiliation{Departamento de Física, Universidad de Santiago de Chile, Av. Victor Jara 3659, Santiago, Chile}
\affiliation{Millennium Nucleus on Young Exoplanets and their Moons (YEMS) }
\affiliation{Center for Interdisciplinary Research in Astrophysics Space Exploration (CIRAS), Universidad de Santiago de Chile, Chile}

\author[0000-0002-7625-1768]{Santiago Orcajo}
\affiliation{Facultad de Ciencias Astronomicas y Geofisicas, Universidad Nacional de La Plata, Paseo del Bosque S/N, 1900 La Plata, Argentina}

\author[0000-0001-8450-3606]{P.H. Nogueira}
\affiliation{Millennium Nucleus on Young Exoplanets and their Moons (YEMS), Chile}

\author[0000-0001-5073-2849]{A.S Hales}
\affiliation{National Radio Astronomy Observatory, 520 Edgemont Road, Charlottesville, VA 22903-2475, United States of America}
%\affiliation{Joint ALMA Observatory, Avenida Alonso de Cordova 3107, Vitacura 7630355, Santiago, Chile}
\affiliation{Millennium Nucleus on Young Exoplanets and their Moons (YEMS), Chile}

\author[0000-0002-1575-680X]{J.M. Miley}
\affiliation{Departamento de Física, Universidad de Santiago de Chile, Av. Victor Jara 3659, Santiago, Chile}
\affiliation{Millennium Nucleus on Young Exoplanets and their Moons (YEMS) }
\affiliation{Center for Interdisciplinary Research in Astrophysics Space Exploration (CIRAS), Universidad de Santiago de Chile, Chile}

\author[0000-0002-3244-1893]{Fernando R. Rannou}
\affiliation{Departamento de Ingenier\'ia Inform\'atica, Universidad de Santiago de Chile, Av. Libertador Bernardo O'Higgins 3363, Estación Central, Santiago}

\author[0000-0002-5903-8316]{Alice Zurlo}
\affiliation{Instituto de Estudios Astrofisicos, Facultad de Ingeniería y Ciencias,
Universidad Diego Portales,  Av. Ejercito 441, Santiago, Chile}
\affiliation{Millennium Nucleus on Young Exoplanets and their Moons (YEMS), Chile}

\collaboration{20}{}

\author{}
\affiliation{}
\affiliation{}

\author{}
\altaffiliation{}
\affiliation{}

\author{}
\affiliation{}
\affiliation{}

\begin{abstract}
    The size of a protoplanetary disk is a fundamental property, yet most remain unresolved, even in nearby star-forming regions (d $\sim$ 140-200 pc). We present the complete continuum size distribution for the $105$ brightest protoplanetary disks (M$_{\text{dust}}$ $\gtrsim$ 2 M$_{\oplus}$) in the Ophiuchus cloud, obtained from ALMA Band 8 (410 GHz) observations at 0.05$^{\prime\prime}$ (7 au) to 0.15$^{\prime\prime}$ (21 au) resolution. This sample includes 54 Class II and 51 Class I and Flat Spectrum sources, providing a comprehensive distribution across evolutionary stages. We measure the Half Width at Half Maximum (HWHM) and the radius encircling $68\%$ of the flux ($R_{68\%}$) for most non-binary disks, yielding the largest flux-limited sample of resolved disks in any star-forming region. The distribution is log-normal with a median value of  $\sim$14 au and a logarithmic  standard deviation $\sigma_{\log} = 0.46$ (factor of 2.9 in linear scale). Disks in close binary systems ($<$ 200 au separation) have smaller radii, with  median value of $\sim$5 au, indicating efficient radial drift as predicted by dust evolution models. The size distribution for young embedded objects (SED Class I and Flat Spectrum, age $\lesssim$ 1 Myr) is similar to that of Class II objects (age $\sim$ a few Myr), implying that pressure bumps must be common at early disk stages to prevent mm-sized particle migration at au scales.

\end{abstract}


\section{Introduction} \label{sec:intro}

The sizes of protoplanetary disks, together with
the distribution of mass within them, establish the basic architectures of planetary systems.
Forming the rocky planets observed by Kepler \citep{2015ApJ...809....8B} requires only a few Earth
masses of dust within $<$ 1 au from the star while forming a Solar System analog requires
significantly more mass and a much larger disk. Therefore, disk sizes and surface density
profiles are the fundamental inputs for population synthesis models \citep{2012A&A...541A..97M, 2017RMxAC..49...76R} trying to understand and reproduce the observed population
of extrasolar planets. However, disk sizes are surprisingly difficult to measure given that 1) disks do not have well-defined edges, disk sizes depend on the tracer (gas vs. dust), and the methods adopted to measure them, and 2) disks seem to be significantly smaller than originally observed.

The first interferometric surveys that were able to measure disk sizes at
(sub)mm wavelengths \citep{1996A&A...309..493D, 2009ApJ...700.1502A,  2009ApJ...701..260I} observed the brightest (sub)mm sources identified by previous single-dish observations \citep{1990AJ.....99..924B,2005AAS...207.3901W} and obtained relatively large disk sizes (r $>$ 50 au). 
However, these early surveys suffered from a severe observational bias since brighter disks tend to be systematically larger than
fainter ones \citep{2021MNRAS.506.5117T,2018ASPC..517..137A,2017ApJ...845...44T}.

Recent ALMA surveys have dramatically increased the
number of disks that have been imaged in nearby molecular clouds at modest resolution (e.g.,  $\sim$30 au;  \cite{2016ApJ...828...46A,
2016ApJ...831..125P, 2016ApJ...827..142B,2019MNRAS.482..698C}). One of the main results from
these surveys has been the fact that most disks remain unresolved at this resolution and must be quite compact (i.e., r $\lesssim$ 15 au). 
Therefore, even after many years of ALMA operations,  the full size distribution of protoplanetary disks in nearby star-forming regions still remains to be established. 
Disk size measurements are particularly scarce in binary systems, typically harboring unresolved, dynamically truncated disks \citep{2020MNRAS.496.5089Z, 2021MNRAS.501.2305Z}.  
Disk size measurements are still scarce in binary systems, typically harboring unresolved, dynamically truncated disks (Zurlo et al. 2020, 2021).  However,   some observational and theoretical results suggest that dust disks in binary systems  might be even smaller than their corresponding truncation radii \citep{2021MNRAS.504.2235Z,2021MNRAS.507.2531Z}

From a theoretical point of view,  the dust continuum disk sizes observed at mm wavelengths are of great interest because, if left unimpeded,  the timescale for the inward migration of 
mm-sized particles in a protoplanetary disk is of the order of 0.1 Myr \citep{2007A&A...469.1169B}. This very short timescale is inconsistent with the observed mm fluxes and sizes of disks that are millions of years old. 
One of the most accepted solutions to prevent the catastrophic loss of mm-sized particles due to their very rapid radial drift is the presence of pressure bumps in the gaseous disk \citep{2012A&A...538A.114P}.  These pressure bumps can be generated by disk instabilities or forming planets \citep{2012A&A...538A.114P, 2020A&A...635A.105P} and can slow down or even halt radial drift completely, trapping dust particles at pressure maxima and regulating the shrinkage of the continuum disk size with time.   
The rings seen in protoplanetary disks when observed at 5-15 au resolution \citep{2018ASPC..517..137A,2018ApJ...869...17L,2021MNRAS.501.2934C}
are similar to the bumps predicted by numerical models and give credence to the idea that these pressure structures play a fundamental role in disk evolution.
In this context, the continuum size of a protoplanetary disk provides direct constraints on the location of the most external pressure bumps in a given system, even when gaps are not spatially resolved by the observations.    

As part of the Ophiuchus DIsk Survey Employing ALMA (ODISEA) project \citep{2019MNRAS.482..698C, 2019ApJ...875L...9W}, here we present ALMA Band 8 (410 GHz)  continuum-only observations at 0.05$^\prime$$^\prime$   (7 au) to 0.15$^\prime$$^\prime$  (21 au) resolution of the brightest $105$ protoplanetary disks in the Ophiuchus Molecular cloud. This sample includes 54 Class II disks and 51 Class I and Flat Spectrum sources.  We resolve every object and investigate the distribution of disk sizes as a function of multiplicity and SED Class.
The observations and data reduction are described in Section 2. In Section 3, we present our main results, which are discussed in Section 4 in the context of disk evolution and planet formation.  

\section{Observation and Data Reduction}

\subsection{Sample Selection and Observations}

Our Band 8   (410 GHz/0.73 mm) sample was selected from the ODISEA survey and is flux-limited. 
We selected all $\sim$100 targets brighter than 4 mJy in Band 6 (230 GHz/1.3 mm) from \cite{2019ApJ...875L...9W}, corresponding to 
M$_{dust}$ $\gtrsim$ 2 M$_{\oplus}$.  
Since $\sim$200 objects were detected by ODISEA in Band 6 (down to a 3-$\sigma$ detection limit of $\sim$0.5 mJy), the Band 8  program is essentially  restricted to the brightest half of all detected objects.  
Given the strong correlation between disk size and disk mass, the Band 8 observations were performed at two different resolutions.  The 45 objects brighter than 20 mJy in Band 6 were observed in Band 8 at 0.15$^\prime$$^\prime$  (21 au)  because they were already known to be resolved at those spatial scales from the lower-resolution Band 6 data. The 55 objects with Band 6 fluxes between 4 and 20 mJy were observed at 0.05$^\prime$$^\prime$  (7 au) angular resolution in Band 8.
Some binary systems were observed at both resolutions because the different components fall in different flux ranges. In the end, a total of 105 disks were observed when counting the individual components in binary systems. 
The full sample of disks is listed in Table \ref{tab: table1}.

The observations at 0.15$^\prime$$^\prime$  resolution were taken during ALMA Cycle 8 under program 2021.1.00378.S  (PI: L. Cieza) between August 4th and 11th, 2022,  and have baselines between 15 m and 1300 m.  
The  0.05$^\prime$$^\prime$  angular resolution data were obtained between May 23rd and 29th, 2023, in ALMA Cycle 9, under program 2022.1.00480.S (PI: L. Cieza) with baselines ranging from  27 m to 3637 m. 
The continuum bandwidth was 7.5 GHz in both programs. 
The precipitable water vapor was around 2.3 mm during the Cycle 8 observations but stayed below 1 mm for the long-baseline observations in Cycle 9.  

\subsubsection{Data Reduction and Analysis} 
The data reduction was performed using the CASA (Common Astronomy Software Applications) \citep{2022PASP..134k4501C} package, following the standard ALMA data processing pipeline. The calibration process included phase and amplitude corrections using observed calibrators, and residual bad data were flagged after initial calibration. Imaging of the data was carried out using the multi-scale CLEAN algorithm. 
We used the Briggs weighting scheme with a robust value of 0.5. 
The final beam sizes for the Cycle 8  and Cycle 9 data were 0.14$^\prime$$^\prime$ $\times$ 0.18$^\prime$$^\prime$ and 0.04$^\prime$$^\prime$ $\times$ 0.06$^\prime$$^\prime$, respectively. We performed auto masking and phase-only self-calibration to all the targets.  

Amplitude self-calibration was attempted but was not implemented because it did not improve the quality of the data.
A typical rms of 0.4-0.5 mJy/beam was achieved.  
All the objects were phase-centered.
An automatic phase centering was performed for the targets without any central cavities, for targets with central cavities a manual phase centering was performed.

Since protoplanetary disks lack well-defined outer edges, there is no universally accepted method for measuring their sizes. 
Measuring the FWHM from a two-dimensional Gaussian fit is one of the simplest approaches.  
This can be done in CASA either in the image plane (via the \textit{imfit} task) or in the visibility plane (via \textit{uvmodelfit}). 
Both methods have some limitations. 
On the one hand, image-plane measurements require deconvolution from the beam and are sensitive to imaging parameters.   On the other hand, visibility fitting becomes challenging for binary systems.
However, we find that, for our sample, both methods agree within $3\%$ across the entire disk size range (see Fig. \ref{App_fig}, in the Appendix). As reliable sizes for binaries are only available in the image plane, we adopt these measurements for our Gaussian fits. 
The Gaussian fit results (coordinates, FWHM of the major and minor axes, position angles, and fluxes) are listed in Table \ref{tab: table1}.
Depending on the signal-to-noise (S/N)  of the observations,  \textit{imfit} can provide deconvolved disk sizes that are even smaller than the beam.  In our case, all the objects observed at 0.15$^\prime$$^\prime$  angular resolutions were well resolved and the S/N of the objects observed at  0.05$^\prime$$^\prime$ ranged from $\sim$10 to $\sim$200, allowing us to measure HWHM values as small as $\sim$0.015$^\prime$$^\prime$.

Another common approach to measure disk sizes is to create a deprojected radial profile from the image plane and calculate the radius that contains $68\%$ of the total flux (R$_{68\%}$), as described in \cite{2017ApJ...845...44T}. 
R$_{68\%}$ can also be estimated by fitting an analytical radial profile to the visibility plane \citep{2020ApJ...895..126H}.
Whenever possible, we obtained radial brightness profiles using the \textit{Frank}  code  \citep{2020fyah.confE..15J} 
% see reference here
%https://github.com/discsim/frank
and the results of the Gaussian fitting as initial parameters. 
%We also employed the Frank profiles to estimate disk sizes, 
We integrated the flux until the radial profile first crossed zero and then calculated the radii enclosing $68\%$ (R$_{68\%}$) and $90\%$ (R$_{90\%}$) of the total flux as alternative size metrics. We note that \textit{Frank} can fail on close binary systems or edge-on disks and when the S/N is low. Overall, we obtained  \textit{Frank} sizes for 80 disks. 
R$_{68\%}$ and R$_{90\%}$\footnote{The R$_{90\%}$ values are not used in our analysis but are provided to allow future comparisons.} values are also listed in Table \ref{tab: table1}. 
A comparison of the R$_{68}$ sizes from the  \textit{Frank} profiles and the HWHM values
from the Gaussian fits (see Fig \ref{App_fig}, right panel, in Appendix) shows a strong correlation, with a best-fit line slope of 0.78, representing a useful conversion factor between the two methods for measuring disk sizes.
This correlation extends to the smaller sources, showing that both IMFIT and Frank are able to resolve even the faintest targets. 
%In Fig. \ref{App_fig-3} in the Appendix, we explore the limits of the the Frank resolution by fitting a point source, and find that our Band 8 observations are indeed sensitive to disks as small as $\sim$2 au in radius.
To verify that even the smallest sources in our sample are partially resolved by the data, in Fig. 4 in the appendix, we plotted the deprojected visibilities as a function of baseline for one of our smallest sources (HWHM = 21 mas) and show that they are not consistent with a point source (which should have flat visibilities).   We also compared the Frank profile to the deconvolved Gaussian size given by   \textit{imfit} for the same source and slightly larger source (Figure 5).  We find that, for well-resolved sources,   \textit{imfit} reproduces well the intensity profile given by Frank.  However,  for very small sources,  \textit{imfit}  might underestimate their size.

\section{Results} \label{sec:results}

\subsection{Continuum Disk Sizes}

In order to investigate the continuum disk sizes in our sample, the angular measurements from Table \ref{tab: table1} were converted to physical sizes in au adopting the distance from 
%\cite{2019ApJ...875L...9W}, which in turn come from 
\cite{2018yCat..36160010G}, when available. For objects without a Gaia distance, a mean distance of 
139.4 pc
(from
\cite{2019ApJ...875L...9W}) 
was assumed. 
The full-size distribution for the 105 disks in the sample is shown in Figure 1 (top-left panel), adopting the Gaussian measurements.
We find that the HWHM values range from  1.7 au to 177 au.  The distribution is log-normal, with a 
median value of 13 au and a logarithmic standard deviation of $\sigma_{\log} = 0.46$ (corresponding to a factor of 2.9 in linear space), see Table \ref{tab: Table2}).

% Define a new column type for wrapping text

\begin{table}[h]
\centering % Ensures the entire table environment is centered
\caption{The complete size distribution for 5 disks}
\makebox[\textwidth]{ % Ensure the table is centered within the text width
\resizebox{\textwidth}{!}{%
\begin{tabular}{l c c c c c c c c c c c}
\toprule
Name  & \multicolumn{1}{c}{RA} & \multicolumn{1}{c}{Dec} & \multicolumn{1}{c}{Dist} & \multicolumn{1}{c}{Maj. Axis} & \multicolumn{1}{c}{Min. Axis} & \multicolumn{1}{c}{Pos. Angle} & \multicolumn{1}{c}{B8 Flux} & \multicolumn{1}{c}{Sep.} & \multicolumn{1}{c}{SED} & \multicolumn{1}{c}{$R_{68\%}$} & \multicolumn{1}{c}{$R_{90\%}$} \\
\cline{2-12}
 ALMA Archive & (deg) & (deg) & (pc)  & (mas)  
 & (mas) &
 (deg) & (mJy) &  (au) & Class & (mas) & (mas) \\
\midrule
ODISEA\_C4\_001 & 245.38306 & -23.02786 & 137.0 & 185.0 $\pm$ 29.0 & 26.0 $\pm$ 19.0 & 167.0 $\pm$ 3.8 & 12.8 $\pm$ 0.3 &  \nodata & II & 173.3 $\pm$ 2.17  & 263.0 $\pm$ 4.50  \\
ODISEA\_C4\_021A & 246.40308 & -24.26180 &142.8 & 53.0 $\pm$ 4.2 & 41.0 $\pm$ 4.5 & 139.0 $\pm$ 17.0 & 20.4 $\pm$ 0.76 & 46.0 & II & \nodata  & \nodata \\ 
ODISEA\_C4\_100 & 246.90966 & -24.61627 & 140.2 & 156.0 $\pm$ 5.0 & 144.0 $\pm$ 3.7 & 109.0 $\pm$ 19.0 & 123.3 $\pm$ 1.5 & \nodata & II & 78.5 $\pm$ 1.3  & 123.4 $\pm$ 2.39  \\
ODISEA\_C4\_117A & 247.18866 & -24.47191 & 81.9 & 385.0 $\pm$ 29.0 & 329.0 $\pm$ 24.0 & 85.0 $\pm$ 33.0 & 224.0 $\pm$ 13.0 & 17.0 & II & 158.3 $\pm$ 6.93  & 213.2 $\pm$ 1.45  \\
ODISEA\_C4\_007 & 245.60530 & -23.49830 & 138.3 & 149.0 $\pm$ 9.5 & 117.0 $\pm$ 9.2 & 9.7 $\pm$ 14.1 & 40.6 $\pm$ 1.4 & \nodata & II & 63.5 $\pm$ 4.94 & 98.5 $\pm$ 7.33  \\
%\label{Table 1}
\midrule
\end{tabular}%
}
}
\footnotesize{Note: For the object ODISEA\_C4\_021A, an R$_{68\%}$ and  R$_{90\%}$ value of \nodata is listed because it is part of a close binary system. R$_{68\%}$ and  R$_{90\%}$ values were measured only for non-binary disks in this table.}
%\caption{The Full sample of the disks}
\label{tab: table1}
\end{table}

\subsection{Size distribution as a function of multiplicity} 

Since stellar companions are expected to truncate each other's disks at 0.3-0.5 times the orbital separation 
%\citep{1977MNRAS.181..441P};\citep {1994ApJ...421..651A}
\citep{1977MNRAS.181..441P,1994ApJ...421..651A},
disks around close binary systems are expected to be significantly smaller than those around single stars (or wide-separation binaries).  
To examine the effect of stellar companions on disk sizes, we divide the sample into two groups:  stars belonging to close-binary systems  (projected sep. $<$ 200 au) and stars lacking close companions.   
The binary systems were identified from two sources, our own ALMA images and the Near-IR Adaptive Optics survey performed by \cite{2020MNRAS.496.5089Z}, which detected binary systems down to $\sim$ 6 au projected separations. Overall, our sample includes a total of 30 disks in 20 binary systems because only in $10$  cases the disks were detected around both components of the system.  Of these 30 disks in multiple systems, 17 belong to close binaries and 13 are part of wide separation systems and are treated as single stars for statistical purposes.

\begin{figure}[hb!]
    \centering
    \begin{minipage}[b]{0.45\textwidth}
        \centering
        \includegraphics[width=\textwidth]{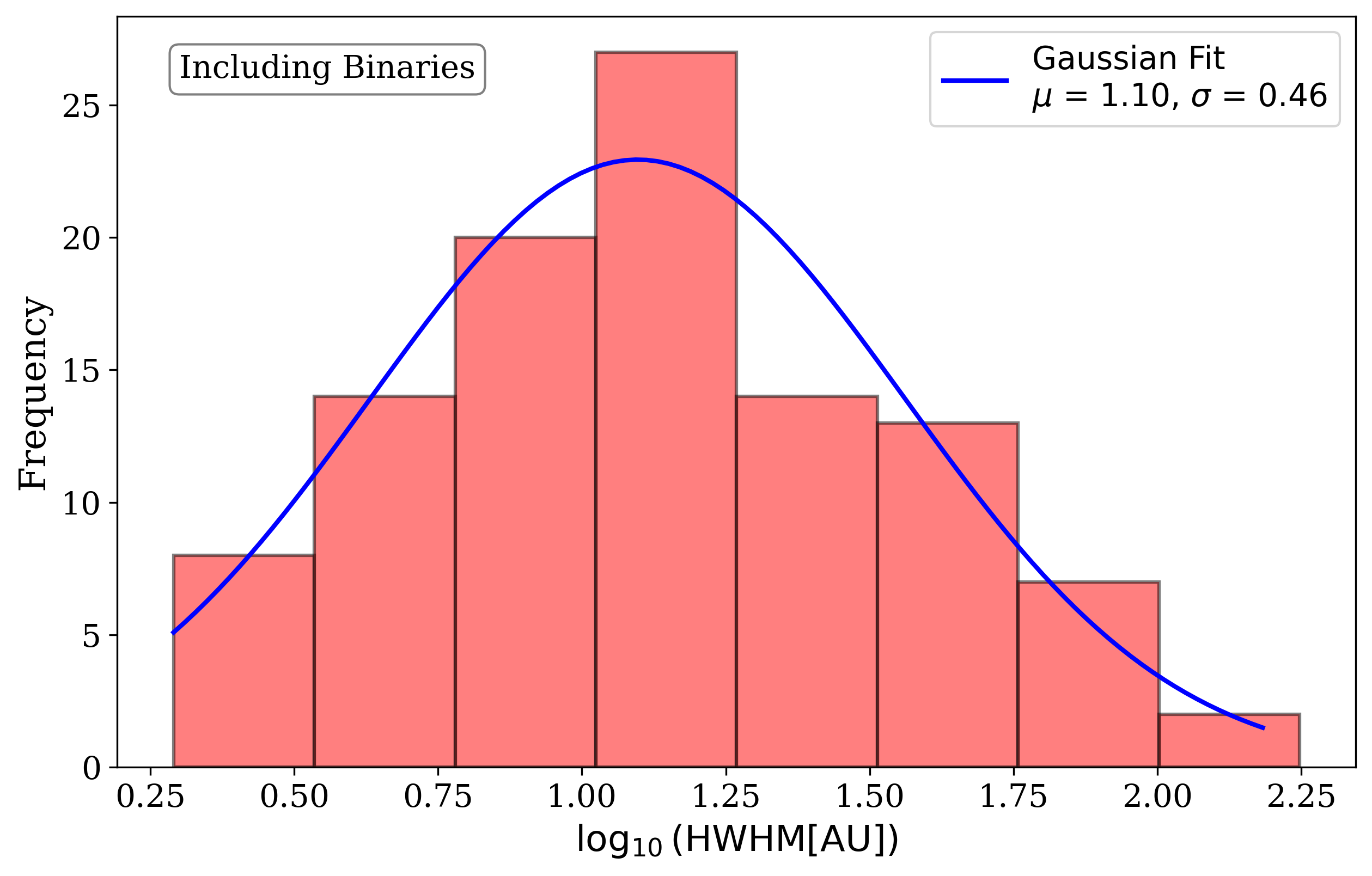}
        %\caption*{Top Left: A.png}
    \end{minipage}
    \hfill
    \begin{minipage}[b]{0.45\textwidth}
        \centering
        \includegraphics[width=\textwidth]{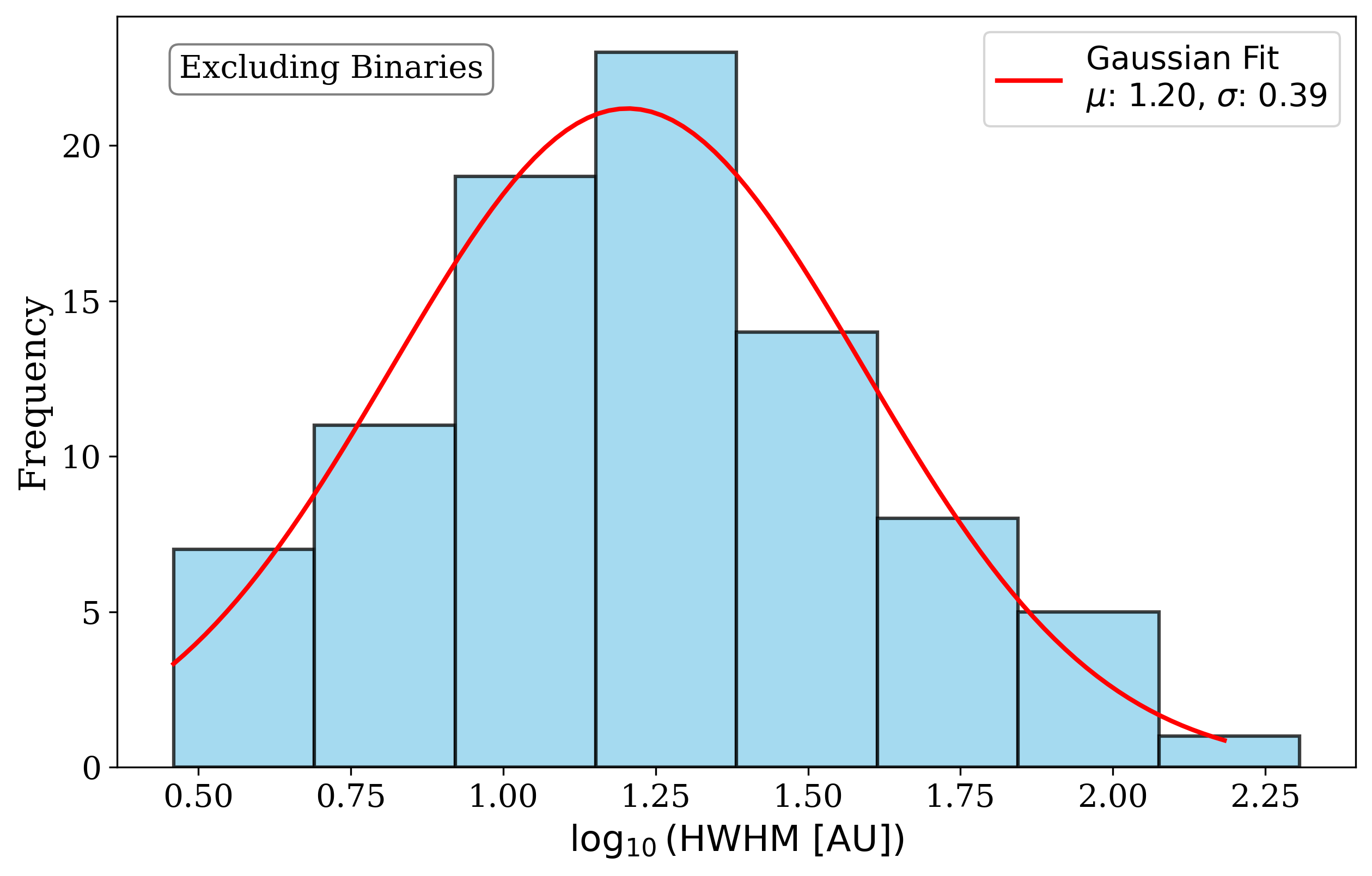}
        %\caption*{Top Right: B.png}
    \end{minipage}

    \vspace{0.5em}

    \begin{minipage}[b]{0.45\textwidth}
        \centering
        \includegraphics[width=\textwidth]{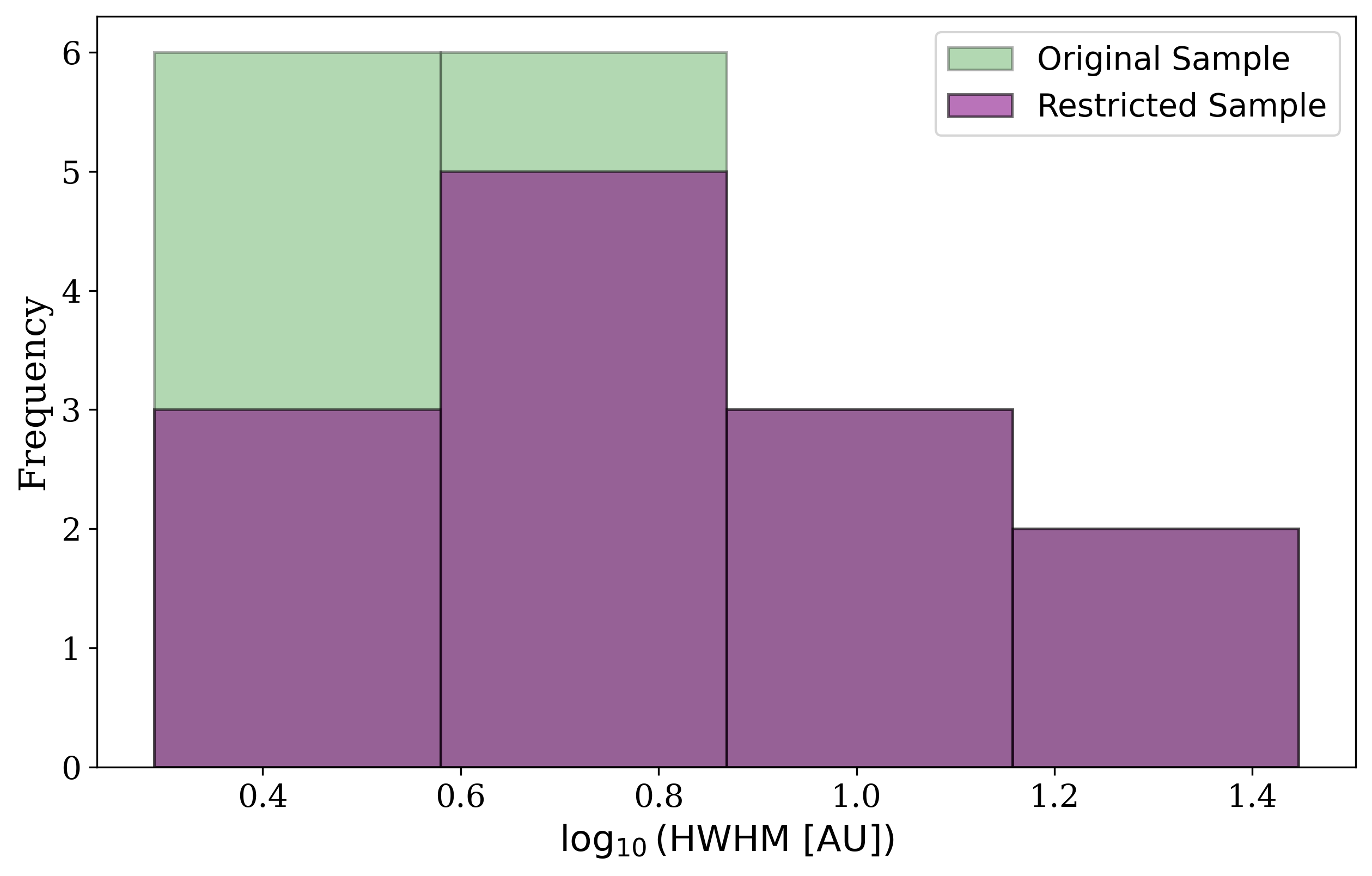}
        %\caption*{Bottom Left: C.png}
    \end{minipage}
    \hfill
    \begin{minipage}[b]{0.45\textwidth}
        \centering
        \includegraphics[width=\textwidth]{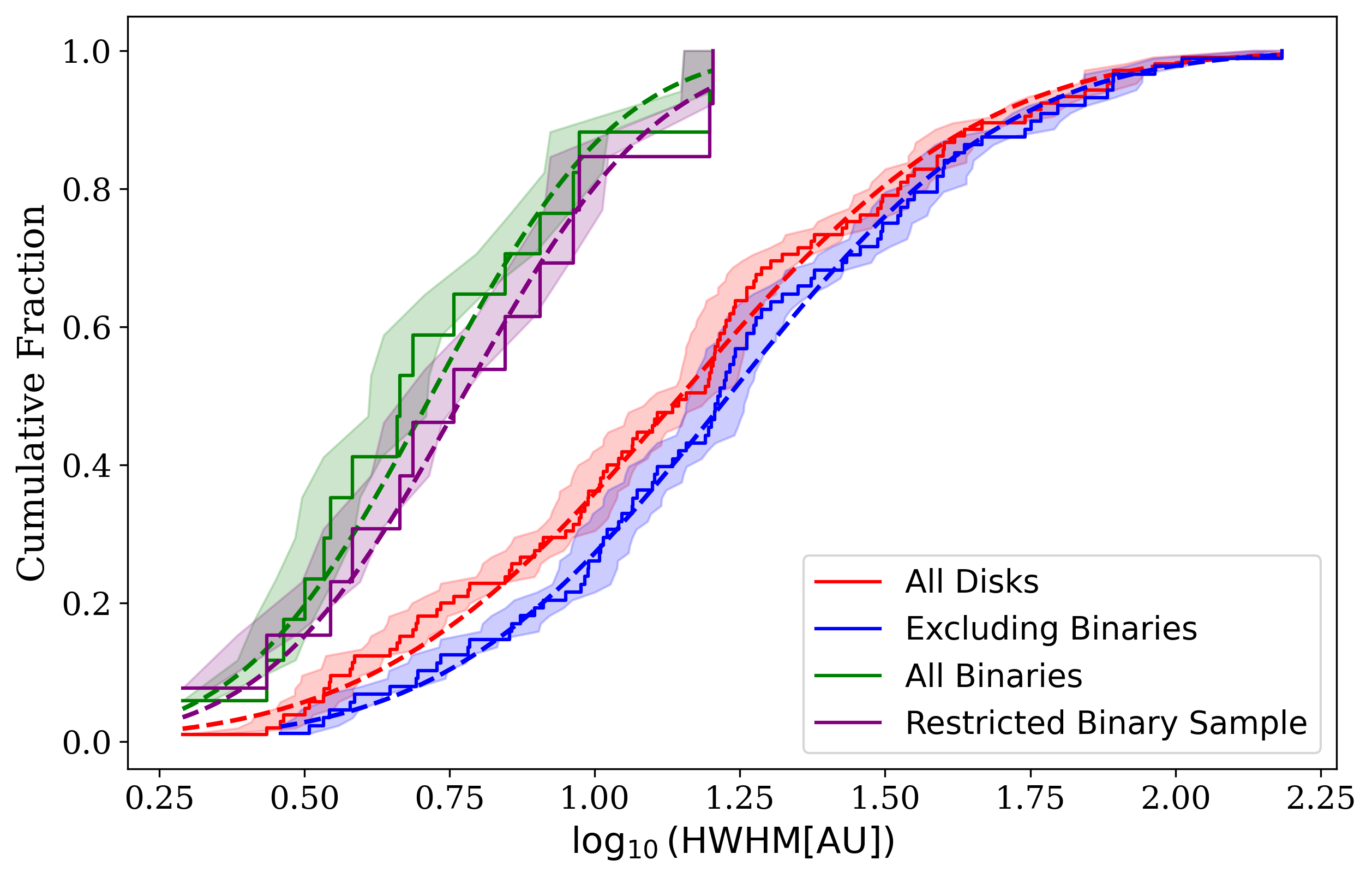}
        %\caption*{Bottom Right: D.png}
    \end{minipage}

    \vspace{0.5em}

    \begin{minipage}[b]{0.45\textwidth}
        \centering
        \includegraphics[width=\textwidth]{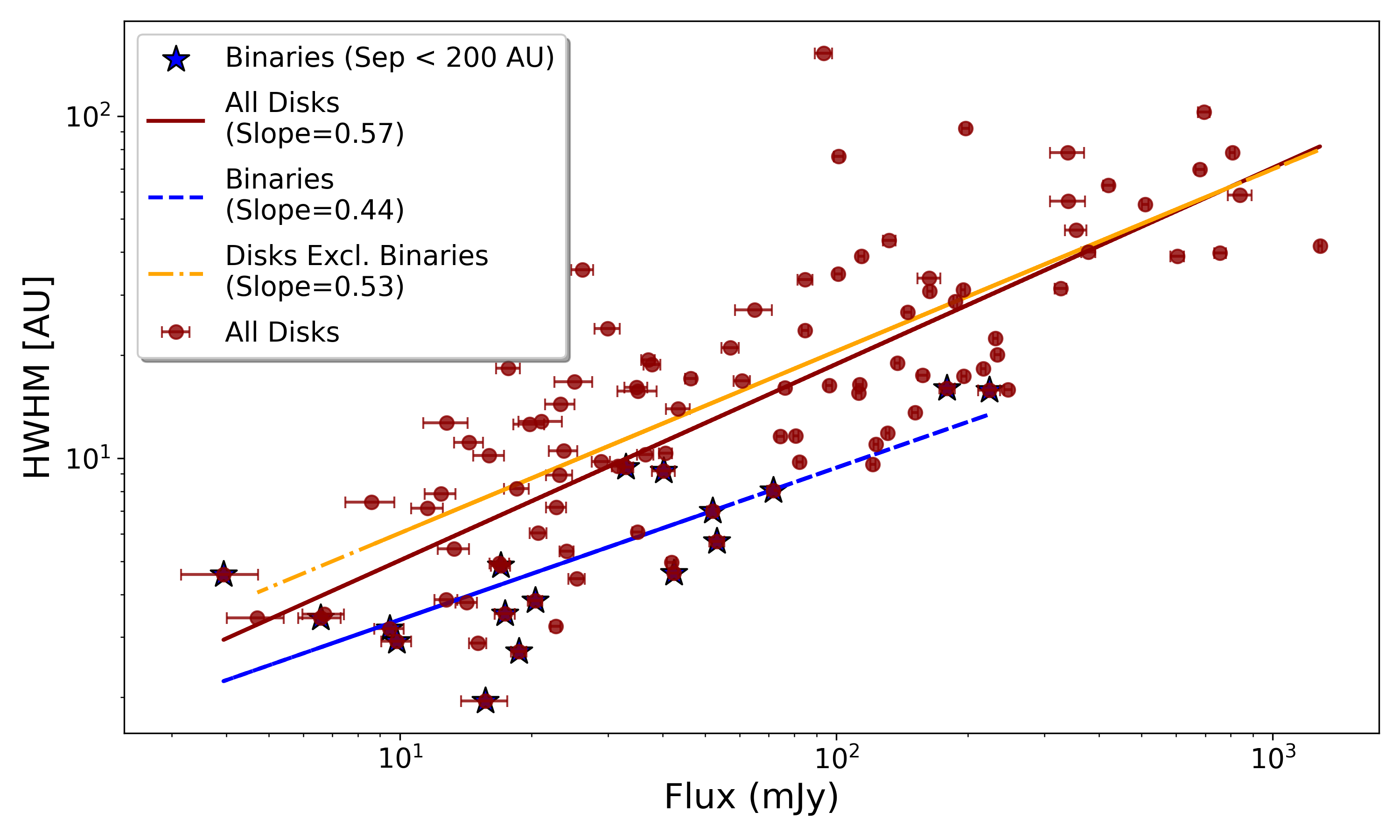}
        %\caption*{Bottom Left: C.png}
    \end{minipage}
    \hfill
    \begin{minipage}[b]{0.45\textwidth}
        \centering
        \includegraphics[width=\textwidth]{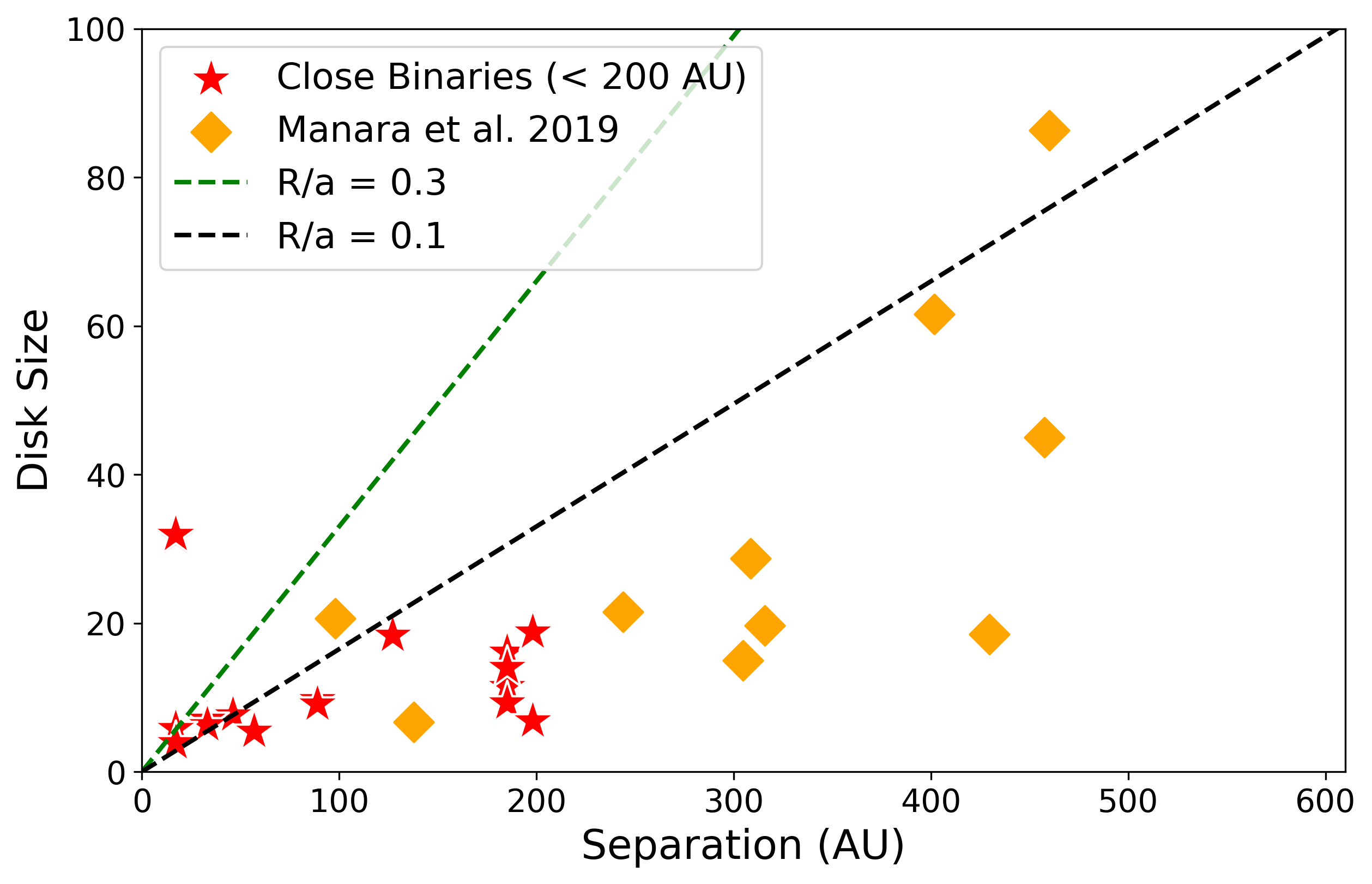}
        %\caption*{Bottom Right: D.png}
    \end{minipage}
    
    \caption{Histograms of the size distribution of the $\sim$100 brightest systems in Ophiuchus (top panels) with Gaussian fits from \texttt{scipy.optimize.curve\_fit} function in Python. The figures show the size distributions including individual components of close binary (sep. $<$ 200 au) systems (left panel) and
excluding them (right panel). 
The middle-left panel corresponds to the disk size distribution of stars only in close binary systems 
and the restricted sample of binary stars (ie., excluding the faintest secondary components).
The middle-right panel presents the cumulative versions of the four distributions. 
The shaded regions correspond to the uncertainties in the Gaussian fits listed in Tale 2. 
In the bottom-left panel, we plot the disk size as a function of  Band 8 flux, finding essentially the same dependence as 
\cite{2020ApJ...895..126H} found for Ophiuchus disks in Band 6 
(Size $\propto$ L$_{mm}^{0.6}$). Close binaries are indicated as labeled.  
The bottom-right panel shows the disk size as a function of binary separations. The 0.1 and 0.3 ratios are indicated.   
The vast majority of objects fall below the 0.1 line. Some outliers are expected because these are  projected  (not physical) separations.  Disk sizes from Manara et al. (2019)  are also shown, but  note that their sizes are R$_{95\%}$ instead of the HWHM measurements we adopted for binary systems.   
} 

    \label{size}
\end{figure}

In Figure \ref{size},  we also show the size distribution of disk sizes excluding disks in close binary systems (top-right panel) and the size distribution for close binaries alone (middle-left panel). 
The cumulative distribution of the full sample and the two sub-samples are plotted in the middle-right panel, together with the Gaussian fits for the observed distributions. 
We find that removing the close binaries has a minor effect on the 
full distribution, increasing the 
the median disk size from  $\sim$14 au to $\sim$16 au
%logarithmic mean from
%$\mu_{\log}$ = 1.20 au to $\mu_{\log}$ = 1.10 au, 
%
and decreasing the standard deviation from $\sigma_{\log}$ = 0.46 to $\sigma_{\log}$ = 0.39.
According to the Kolmogorov-Smirnov (K-S) test,  the distributions of disk sizes including or excluding close binaries are very similar (see Table \ref{tab: table3}).  
On the other hand, the size distribution of close binaries alone is clearly different with a median disk size of 4.6 au
%, with a log-normal, with a logarithmic mean of $\mu_{\log} = 0.71 \, \mathrm{au}$ (corresponding to $5.1 \, \mathrm{au}$ in linear space) 
and a logarithmic standard deviation of $\sigma_{\log} = 0.25$ (corresponding to a factor of 1.8 in linear space).  
We note that the binary systems include secondary disks that do not necessarily meet the flux requirements in our flux-limited sample or that were only detected in the Band 8 high-resolution observations. To correct for this bias, we define a “restricted binary sample” which excludes 4 secondary disks fainter than 4 mJy x [B8/B6], where   [B8/B6] is equal to 3.34, the mean Band 8 to Band 6 ratio of the entire sample.  This restricted binary sample is also shown in Fig. 1 (middle panels) and included in Table 2.
As shown in Table 3, the K-S test confirms a significant difference with respect to the sample excluding close binaries, although with a slightly lower significance. 
We note that the results remain essentially the same if separation thresholds of 100 au or 300 au are used to define close binary systems, instead of the 200 au we adopted.

Another potential bias when comparing disk sizes is the difference in the underlying distributions of host stellar masses between single stars and binaries,  given the known correlations between disk luminosity and stellar mass \cite{2019A&A...631L...2M} and between disk luminosity and disk sizes \citep{2020ApJ...895..126H}.   In particular,  secondary objects in binary systems are likely to be biased toward lower host masses and disk luminosities and hence toward smaller disk sizes. To disentangle these dependencies,  we plot the disk sizes as a function of flux for all disks and close-binary systems (See Fig. 1, bottom-left panel). We find that disks in binary systems indeed span a smaller range of disk fluxes.  However, it is also clear that for a given mm flux, disks in binary systems are systematically smaller.
 
\clearpage

\begin{table}[h]
\centering
\caption{Size and Flux Statistics for Different Samples}\label{tab: Table2}
\begin{tabular}{|l|c|c|}
\hline
\textbf{Sample} & \textbf{Median ($\mu$)} & \textbf{Standard Deviation ($\sigma$)} \\
               &      (au) &          Log.          (au)                    \\
\hline
\multicolumn{3}{|c|}{\textbf{Whole sample and binaries}} \\
\hline
Whole Sample & $14.37 \pm 1.38$ & $0.46 \pm 0.02$ \\
Excluding Binaries & $16.37 \pm 1.57$ & $0.39 \pm 0.02$ \\
Binaries Only & $4.62 \pm 0.68$ & $0.25 \pm 0.04$ \\
Restricted Sample & $5.88 \pm 1.08$ & $0.26 \pm 0.05$ \\
\hline
\multicolumn{3}{|c|}{\textbf{SED Classes (Including close binaries)}} \\
\hline
Class I \& F & $14.37 \pm 1.75$ & $0.41 \pm 0.04$ \\
Class II & $14.70 \pm 1.02$ & $0.39 \pm 0.04$ \\
\hline
\multicolumn{3}{|c|}{\textbf{SED Classes (Excluding close binaries)}} \\
\hline
Class I \& F & $15.84 \pm 1.52$ & $0.36 \pm 0.04$ \\
Class II & $16.21 \pm 1.97$ & $0.38 \pm 0.04$ \\
\hline
\multicolumn{3}{|c|}{\textbf{SED Classes (Frank Sizes)}} \\
\hline
Class I \& F  & $14.12 \pm 1.72$ & $0.31 \pm 0.05$ \\
Class II  & $16.98 \pm 1.63$ & $0.31 \pm 0.04$ \\
\hline
\multicolumn{3}{|c|}{\textbf{SED Classes (Fluxes)}} \\
\hline
Class I \& F  & $57.54 \pm 27.56$ & $0.52 \pm 0.05$ \\
Class II  & $83.17 \pm 45.64$ & $0.52 \pm 0.05$ \\
\hline
\end{tabular}
\end{table}

\begin{longtable}{l l l l l}

\caption{Kolmogorov-Smirnov Test Comparison} \label{tab: table3}\\
\toprule
Size Comparison & Sample Sizes & D-value & p-value & Figure Panels\\
\midrule
\endhead
\midrule
\multicolumn{4}{r}{\textit{Continued on next page}} \\
\endfoot
\bottomrule
\endlastfoot
Whole Sample vs. Excluding Close Binaries & (105, 88) & 0.10 & 0.58 & Fig. 1 (Middle Right) \\
Excluding Binaries vs. Close Binaries & (88,17) & 0.66 & $1.0 \times 10^{-6}$ & Fig. 1 (Middle right) \\
Excluding Binaries vs. Restricted Sample & (88,13) & 0.63 & $7.0 \times 10^{-5}$ & Fig. 1 (Middle right) \\
Class I + Flat vs. Class II (incl. close binaries) & (51,54)  & 0.16 & 0.45 & Fig. 2 (Top Right) \\
Class I + Flat vs. Class II (excl. close binaries) & (46,42) &  0.22 & 0.16 & Fig. 2 (Middle Right) \\
Class I + Flat vs. Class II (Frank Sizes) & (42,38)  & 0.12 & 0.87 & Fig. 2 (Bottom right) \\
\toprule
Flux Comparison & Sample Sizes & D-value & p-value & Figure Panels\\
\midrule
Class I + Flat  vs. Class II (incl. close binaries) & (51,54)  & 0.20 & 0.28 & Fig. 4 (Appendix) 
\\
\toprule
Inclination Comparison & Sample Sizes & D-value & p-value & Figure Panels\\
\midrule
Class I + Flat vs. Isotropic Distribution & (51, 1000)  & 0.13 & 0.42 & Fig. 4 (Appendix) 
\end{longtable}
\FloatBarrier

% Updated figure with four panels

%\newpage

\subsection{Size distribution as a function of SED Class}\label{size_sed}

The Ophiuchus molecular cloud is characterized by a significant population of embedded stars (Class I and Flat Spectrum sources), with an estimated age of  $\lesssim$ 0.5-1 Myr,  and mostly located in the L1688 cluster \citep{2009ApJS..181..321E}.  
It also contained a more distributed population of Class II and Class III sources sources with a significant age spread, with stars as old as 6 Myr according to \cite{2020yCat..51590282E}.   We note however that most of the embedded and Class II sources in our sample are associated with the L1688 Cluster.

\begin{figure}[ht!]
    \centering
    \begin{minipage}[b]{0.45\textwidth}
        \centering
        \includegraphics[width=\textwidth]{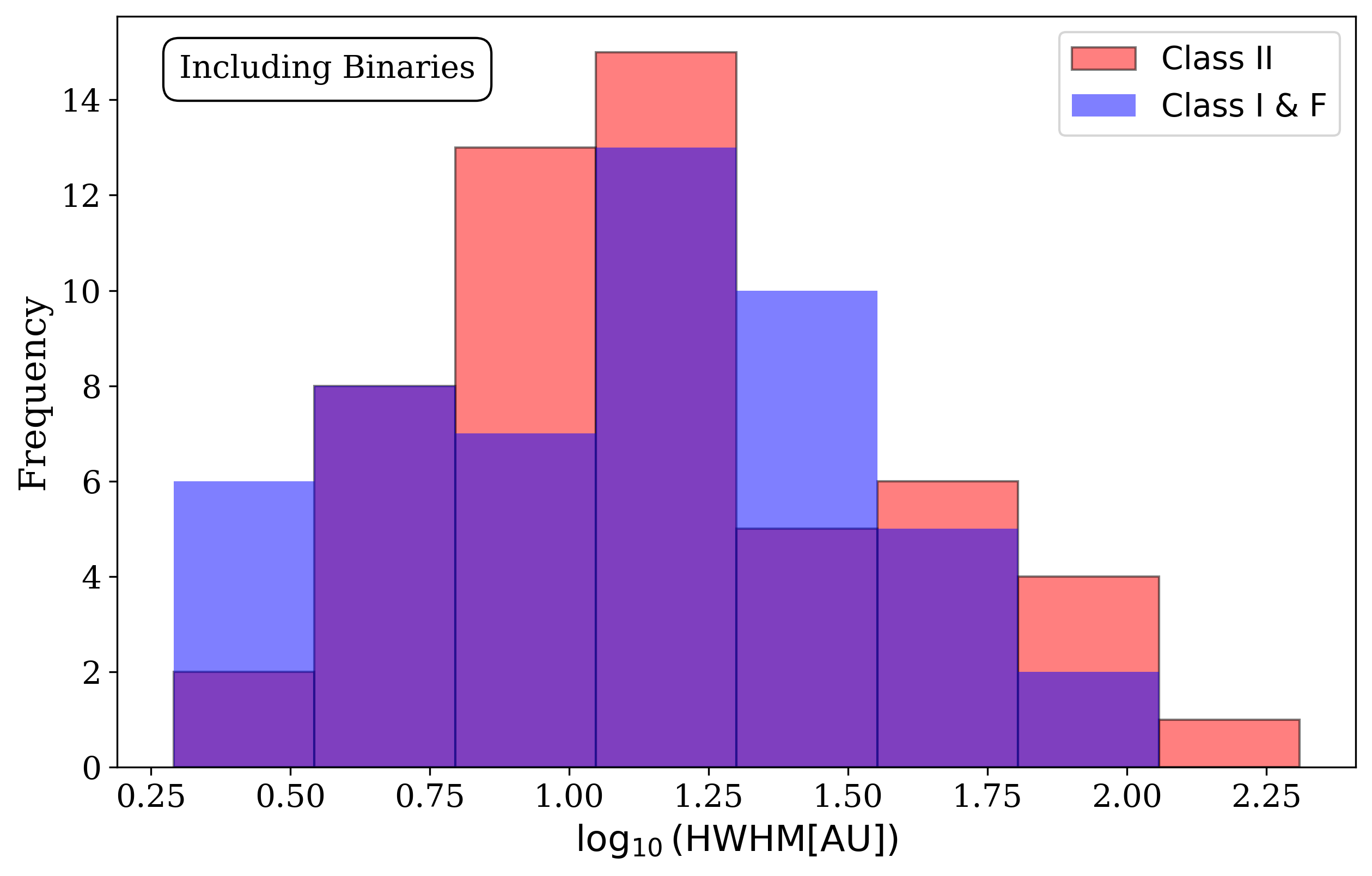}
        %\caption*{Top Left: A.png}
    \end{minipage}
    \hfill
    \begin{minipage}[b]{0.45\textwidth}
        \centering
        \includegraphics[width=\textwidth]{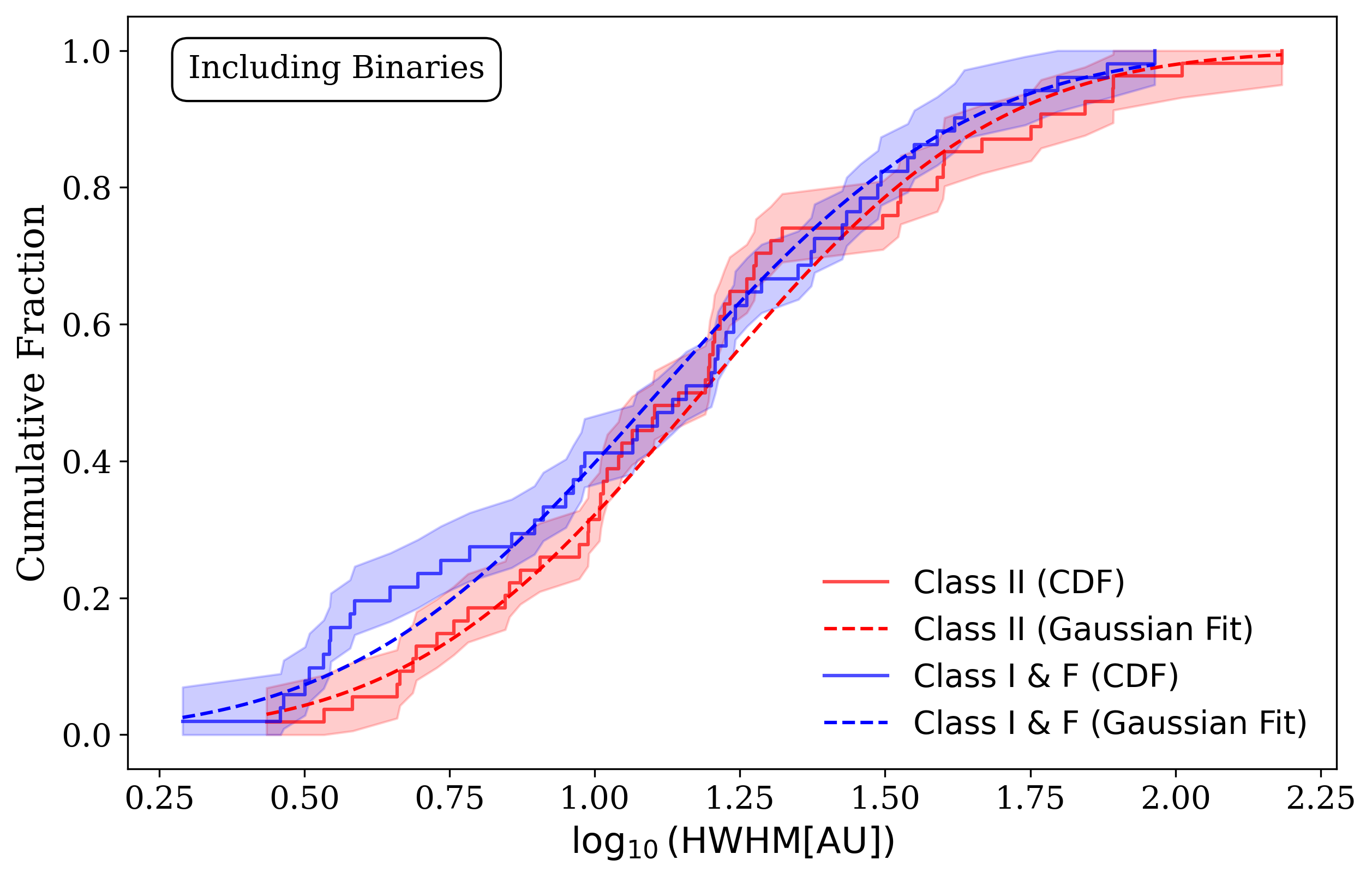}
        %\caption*{Top Right: B.png}
    \end{minipage}

    \vspace{0.5em}

    \begin{minipage}[b]{0.45\textwidth}
        \centering
        \includegraphics[width=\textwidth]{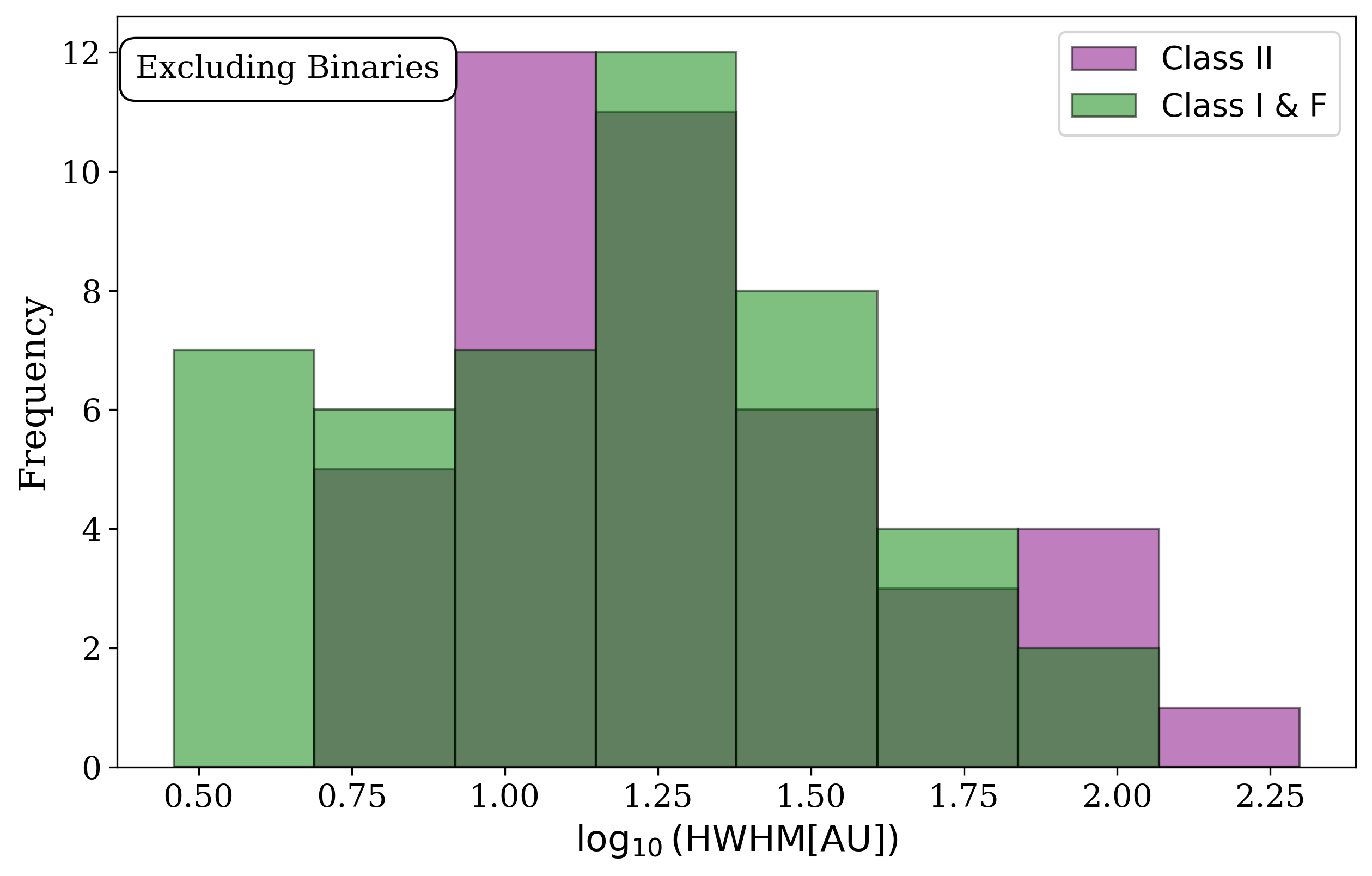}
        %\caption*{Bottom Left: C.png}
    \end{minipage}
    \hfill
    \begin{minipage}[b]{0.45\textwidth}
        \centering
        \includegraphics[width=\textwidth]{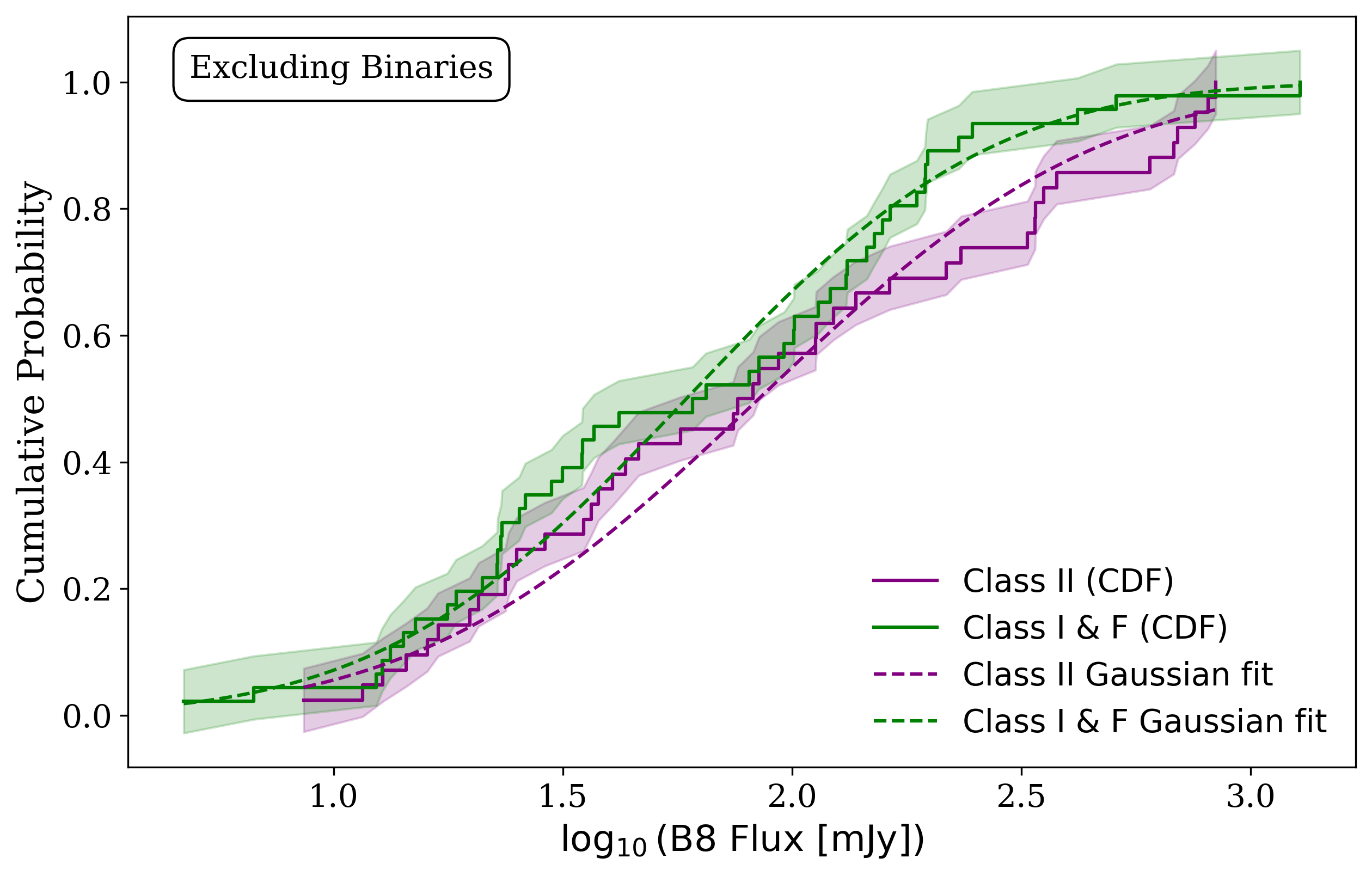}
        %\caption*{Bottom Right: D.png}
    \end{minipage}
    \vspace{0.5em}

    \begin{minipage}[b]{0.45\textwidth}
        \centering
        \includegraphics[width=\textwidth]{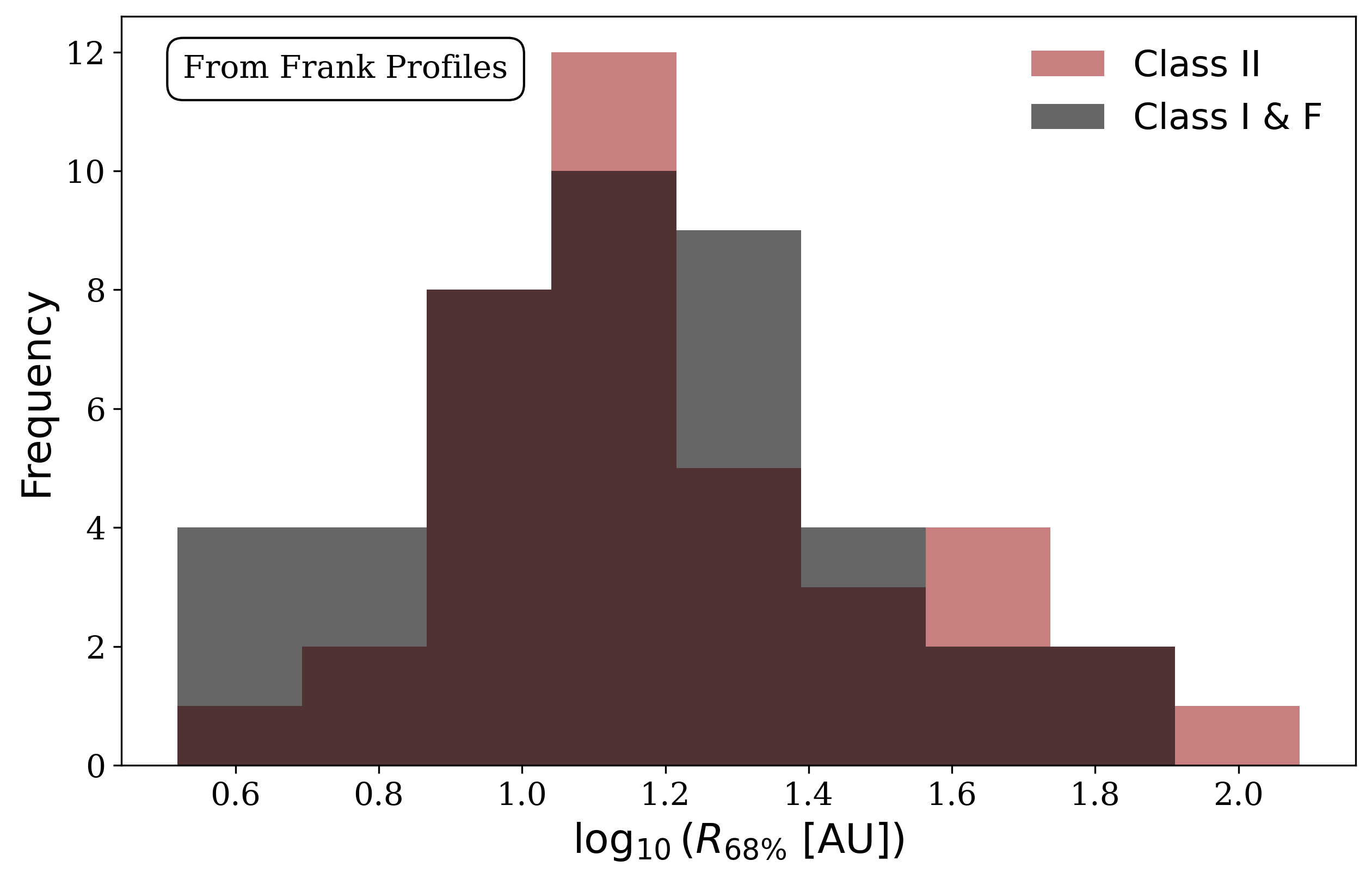}
        %\caption*{Bottom Left: C.png}
    \end{minipage}
    \hfill
    \begin{minipage}[b]{0.45\textwidth}
        \centering
        \includegraphics[width=\textwidth]{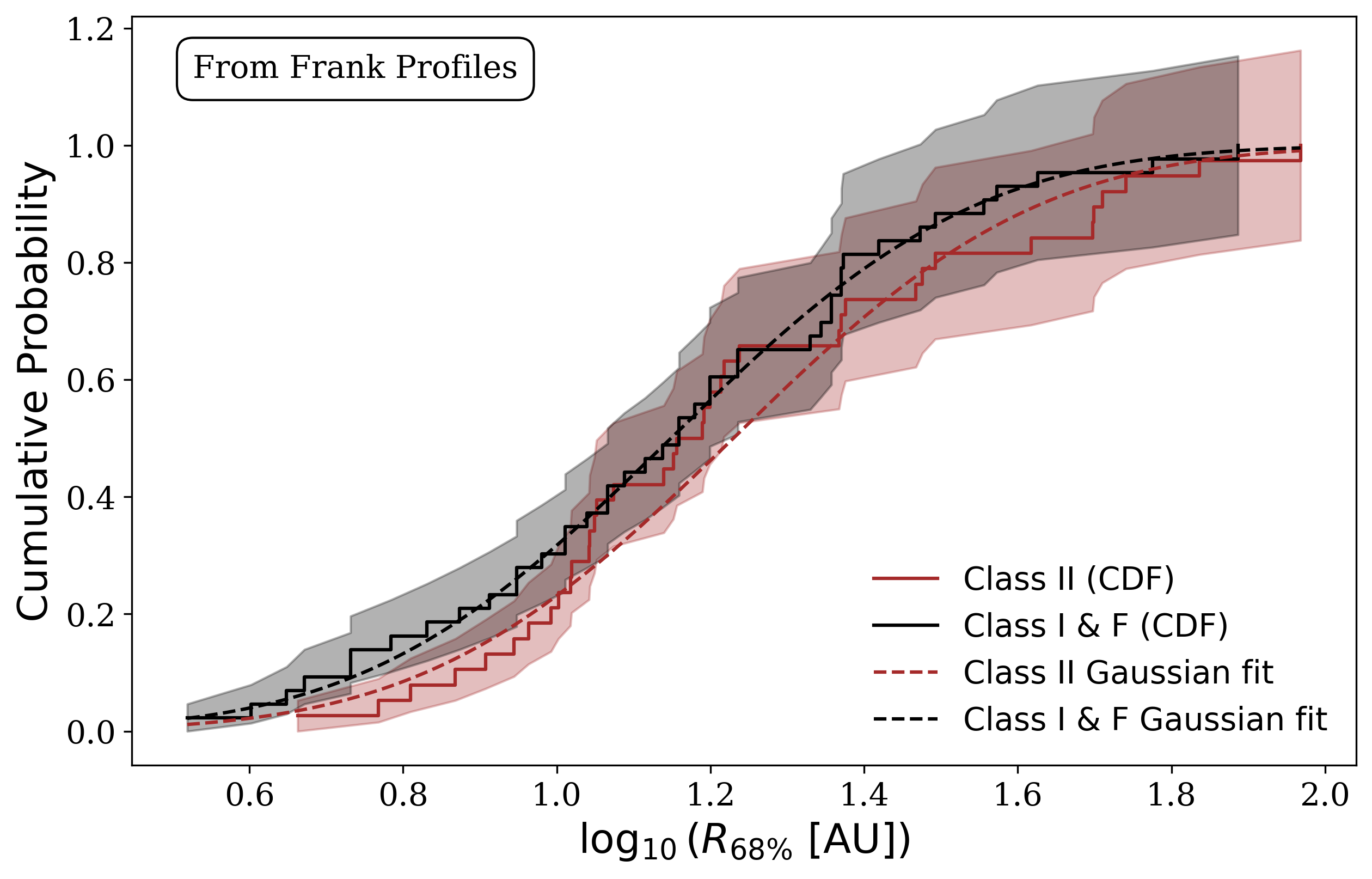}
        %\caption*{Bottom Right: D.png}
    \end{minipage}
    \caption{Histograms of the size distribution of embedded targets (Class I and Flat Spectrum sources) and Class II objects (left panels).
    The top and middle panels correspond to Gaussian size measurements,  
    including and excluding close binaries. 
    The bottom panel shows the size distribution for the 80 disks with Frank Profiles (only available for non-binary disks). 
    The corresponding cumulative distributions are shown on the right panels. 
    Results of the K-S tests show that the distributions of different SED classes are statistically indistinguishable from each other.  
    }
    \label{Class}
\end{figure}

In order to investigate how continuum disk sizes evolve with time,  we conducted a size distribution analysis for both groups (embedded stars vs. Class II sources).
Figure \ref{Class} presents the size distribution of the two SED groups using the Gaussian measurements including close binaries and excluding them (top-left and middle-left panels, respectively). 
The bottom-left panel corresponds to the size distribution adopting the R$_{68\%}$ values from \textit{Frank} (only available for non-binary disks).

We further plot cumulative distribution functions for all cases with their corresponding Gaussian fits (panels on the right). As with the previous analysis, we performed K-S tests to compare the distributions between the two SED groups.
The results of the K-S test are shown in Table \ref{tab: table3}.
These tests show that the distributions remain statistically indistinguishable from each other, regardless of the inclusion or exclusion of the binary systems, or whether we adopt Gaussian sizes or R$_{68\%}$ values from \textit{Frank} profiles. 
Our results indicate that disk sizes are \textit{not} becoming smaller between the embedded stage and the Class II phase as expected from the radial drift of mm-sized particles.   In fact, at face value, Class II disks are slightly larger than disks around embedded sources (see Table 2).  Given the strong dependence of disk size on mm flux, this result would be expected if the Class II disks in our sample were significantly brighter 
than the embedded disks. However, we have verified that this is not the case as we find that the flux distributions of the two samples are indistinguishable from each other according to the K-S test (see Table 3 and left of Figure 4 in the Appendix).  
Another potential bias that we investigated is the possibility that the sample of embedded sources is contaminated by more evolved, but highly inclined, Class II objects \citep{2023ApJ...951....8O}. To test for this possibility we compare the observed distribution of inclinations to that expected for an isotropic distribution of disk orientations (see Figure 4, right panel in the appendix).  However,  we find that the observed inclination distribution of embedded sources is indistinguishable from random orientations (see K-S in Table 3) and conclude that our sample of embedded sources is not significantly contaminated by highly-inclined older objects.
%

%\FloatBarrier

% Updated figure with four panels

%\newpage
\section{Discussion}

\subsection{Impact of Multiplicity on Planetary Architectures} 

In our analysis of the $\sim$100 brightest protoplanetary disks in Ophiuchus, we examined how multiplicity influences size distribution. 
In our flux-limited sample, we find that the stars in close binary systems (sep. $<$ 200 au) have disks that are, on average, more than a factor of two smaller than disks around single stars or wide binary systems (HWHM $\sim$ 5 vs $\sim$ 13 au).
The most straightforward implication is that the formation of Uranus and Neptune analogs should be strongly inhibited in close binary systems, which should be prone to producing compact planetary configurations.
However, models by \cite{2021MNRAS.507.2531Z} recently suggested that stellar companions might also affect the formation of planets in additional ways.  In particular, they find that dust grains should undergo more efficient radial drift in close binaries with respect to wide binaries and single stars.  
This would very quickly ($<$ 1 Myr) reduce the amount of solids available for the formation of the cores of gas giants unless disk substructures can trap the dust and prevent its inward migration at the very early stages of disk evolution. 

The bottom-left panel of Figure 1 shows that, for a given flux (a proxy for dust mass), disks in close binary tend to be smaller than the rest. This is consistent with a more efficient radial drift and supports the notion that dust evolution, and not only tidal truncation, play a role on the continuum disk sizes in binary systems. 
This idea can be tested by comparing binary separations to the expected truncation radius.  The truncation radii are expected to be 0.3-0.5 $\times$ the orbital separations \citep{1997MNRAS.284..821P, 1994ApJ...421..651A}.
Instead, we find that the continuum radii are $<$ 0.1 $\times$ the projected orbital separations (see Figure 1, bottom-right panel), 
confirming the result found by \cite{2019A&A...631L...2M}  for Taurus binaries. 
The tidal truncation radius depends strongly on eccentricity, but \cite{2019A&A...631L...2M} already demonstrated that the eccentricity values would need to be unrealistically high to explain the observed disk sizes by tidal truncation alone. 

The observational results presented herein should allow more detailed comparisons to numerical models as those of \cite{2021MNRAS.507.2531Z}, which can in principle explain disks smaller than the truncation radii in terms of grain growth and radial drift in binary systems. They compared the observed disk sizes in binary systems in Taurus (from \cite{2019A&A...631L...2M}) and Ophiuchus (from \cite{2017ApJ...851...83C}) and obtained good agreement with their models \citep{2021MNRAS.504.2235Z}, but concluded that more resolved disks were needed to derive more robust conclusions.   We note that our results in Ophiuchus are highly complementary to those studied by  \cite{2021MNRAS.507.2531Z} because our disk sizes come from binaries with projected separations smaller than 200 au, while resolved disk from \cite{2019A&A...631L...2M}  and \cite{2017ApJ...851...83C} correspond to disks in binaries with separations larger than 200 au (see Figure 1, bottom-right panel).\footnote{
With the exception of Elias 2-33 (a.k.a ODISEA$\_$C4$\_$94), all the disks in binaries  were known and had ALMA observations (e.g., from Cox et al. 2017 and Zurlo et al. 2021). However, most disks in close binaries were previously unresolved and lacked disk measurements.
}

Detailed comparisons of disk sizes to the outer architectures of extrasolar planetary systems are also highly desirable, but  
our ability to detect extrasolar planets at separations beyond 
$\sim$5 au remains very limited, especially around binary systems.  
For instance, the detection of Uranus and Neptune analogos in extrasolar binary systems is currently restricted to the microlensing technique and few examples exist e.g., \cite{ 2014ApJ...795...42P}. However, the Nancy Grace Roman Space Telescope will dramatically change this situation \citep{2019ApJS..241....3P} and should provide statistics on exoplanets at 1-30 au separations around binary systems that could be compared to disk demographic results.  

\subsection{The Size Distribution Across SED Classes: implications for disk evolution and planet formation}

The lack of evolution in continuum disk sizes between embedded objects and Class II sources found in Section \ref{size_sed} is very interesting in the context of pressure bumps, grain growth and radial drift.   
Pressure bumps can trap dust particles, preventing inward drift and enabling growth into larger bodies \citep{2012A&A...538A.114P, 2020A&A...635A.105P}.
These pressure bumps are consistent with the rings around most of the massive Class II disks observed at high resolutions \citep{2018ASPC..517..137A, 2018ApJ...869...17L, 2021MNRAS.501.2934C} and with the relatively slow evolution of disk sizes seen between 1–10 Myr old systems \citep{2020ApJ...895..126H}.  
Our new ALMA observations includes younger, less massive, and smaller disks than the previous studies mentioned above and suggest that these bumps are ubiquitous and must be present from the early  stages of disk evolution (age $\lesssim$ 1 Myr).
Additional observational evidence supports this idea. For instance, 
our Band 8 data has enough resolution and signal to noise to show gaps and rings in several disks that were previously considered smooth (Bhowmik et. in prep). 
The nearby disk around TW Hydra also shows gaps at au scales \citep{2016ApJ...820L..40A}, and even objects at the deuterium fusion limit can show  substructures when observed at sufficiently high-enough resolution, e.g., 
%\cite{2023MNRAS.526.1545C};  
\cite{2020ApJ...902L..33G}.  

Future observations with ALMA and the ngVLA \citep{2024ApJ...965..110W} at even higher angular resolution should be able to search for substructures in all the disks that remain smooth in our study.    
Meanwhile, the observed continuum size distribution can already provide constraints on the locations of the pressure bumps.
As shown by \cite{2019MNRAS.486.4829R},  disk sizes derived from ALMA continuum observations mostly trace the radius up to which the disks retain grains large enough to emit significantly at (sub)millimeter wavelengths (e.g.,  the locations of the outermost dust traps).
Furthermore, if dust traps are assumed to be caused by the formation of planets (e.g. \cite{2018ApJ...869L..47Z}, numerical models that combine viscous evolution, planet-disk interactions,  grain growth, and radial drift can be used to investigate the observed (lack of) evolution in disk sizes.  The output of such models can be postprocessed with radiative transfer to allow for direct comparison to ALMA observations. Such comparisons can then constrain the underlying populations of planets (in terms of masses and semi-major axes) that are needed to reproduce the observed size distribution in our sample of Ophiuchus disks (Orcajo et al. in prep.), bridging the gap between disk demographics and planet population studies.

It is important to consider that our sample excludes all disks with dust masses $\lesssim$ 2 M$_{\oplus}$.  Such disks might have a different population of planets (e.g., mostly terrestrial planets within $\lesssim$ 1 au).  Therefore, to obtain a more complete picture and to test the idea that dust disk sizes are mostly regulated by planet formation that halts dust drift, it would be very beneficial to explore the full size distribution of disk in nearby molecular clouds. This is within current capabilities, as ALMA can reach a resolution of  $\sim$1 au  at 140 pc using its most extended configuration in Band 8.

The measurement of disk sizes as a function of age can also provide valuable constraints on disk evolution models. In particular, viscous models predict that the disk will expand with time, while models of disks driven by magnetic winds find that the disk sizes should remain close to constant \citep{2022MNRAS.512L..74T}.  However, most continuum ALMA surveys, including ours, lack the sensitivity to detect the faint emission from the outer disk (arising from small dust) to distinguish between the scenarios \citep{2022MNRAS.514.1088Z}. Measuring the sizes of the gaseous disk might be a more direct way to distinguish between viscous and magnetic wind models. \cite{2023ApJ...954...41T} recently found that gas disk sizes are significantly smaller in Upper Sco (age $>$ 5 Myr) than in Lupus and Taurus (Age $<$ 3 Myr). This seems to contradict both viscous and wind-driven evolution and suggests that external photoevaporation could also play an important role on the evolution of disk sizes. 
Investigating disk sizes as a function of time is one of the main goals of the AGE-PRO ALMA Large Program (ALMA Survey of Gas Evolution of PROtoplanetary Disks; Zhang et al. submitted).  Thanks to very deep continuum and molecular line observations,  AGE-PRO will help to constrain the importance of viscous spreading, magnetic winds and photoevaporation in disk evolution (Tabone et al. 2024, submitted; Anania et al. 2024,  submitted).

\begin{acknowledgments}

We thank the anonymous referee for the precious comments, which have helped improve the paper.
A.D. was supported by the Joint China–Chile Committee fund and from  the Millennium Nucleus on Young Exoplanets and their Moons (YEMS), ANID - Center Code NCN2021\_080 and NCN2024\_001..  
T.B. acknowledges financial support from the FONDECYT postdoctorado project number 3230470
L.A.C. and G.B-F. acknowledge support from ANID, FONDECYT Regular grant number 1241056.
L.A.C., T.B., C.G-R., P.C.,
S.C., S.P., F.R., and and A.Z. acknowledge support from the Millennium Nucleus on Young Exoplanets and their Moons (YEMS), ANID - NCN2021\_080 and NCN2024\_001. 
This paper makes use of the following ALMA data: ADS/JAO.ALMA \# 2021.1.00378.S and \# 2022.1.00480. 
ALMA is a partnership of ESO (representing its member states), NSF (USA),
and NINS (Japan), together with NRC (Canada), NSC and ASIAA (Taiwan), and KASI (Republic of Korea), in cooperation with the Republic of Chile. The Joint ALMA Observatory is operated by ESO, AUI/NRAO, and NAOJ. The
National Radio Astronomy Observatory is a facility of the National Science Foundation operated under cooperative agreement by Associated Universities, Inc.

\end{acknowledgments}

\vspace{5mm}
\facilities{ALMA}

\software{CASA}

%\appendix

\section{Additional Figures}

\begin{figure}[h]
    \centering
    \begin{minipage}[b]{0.45\textwidth}
        \centering
        \includegraphics[width=\textwidth]{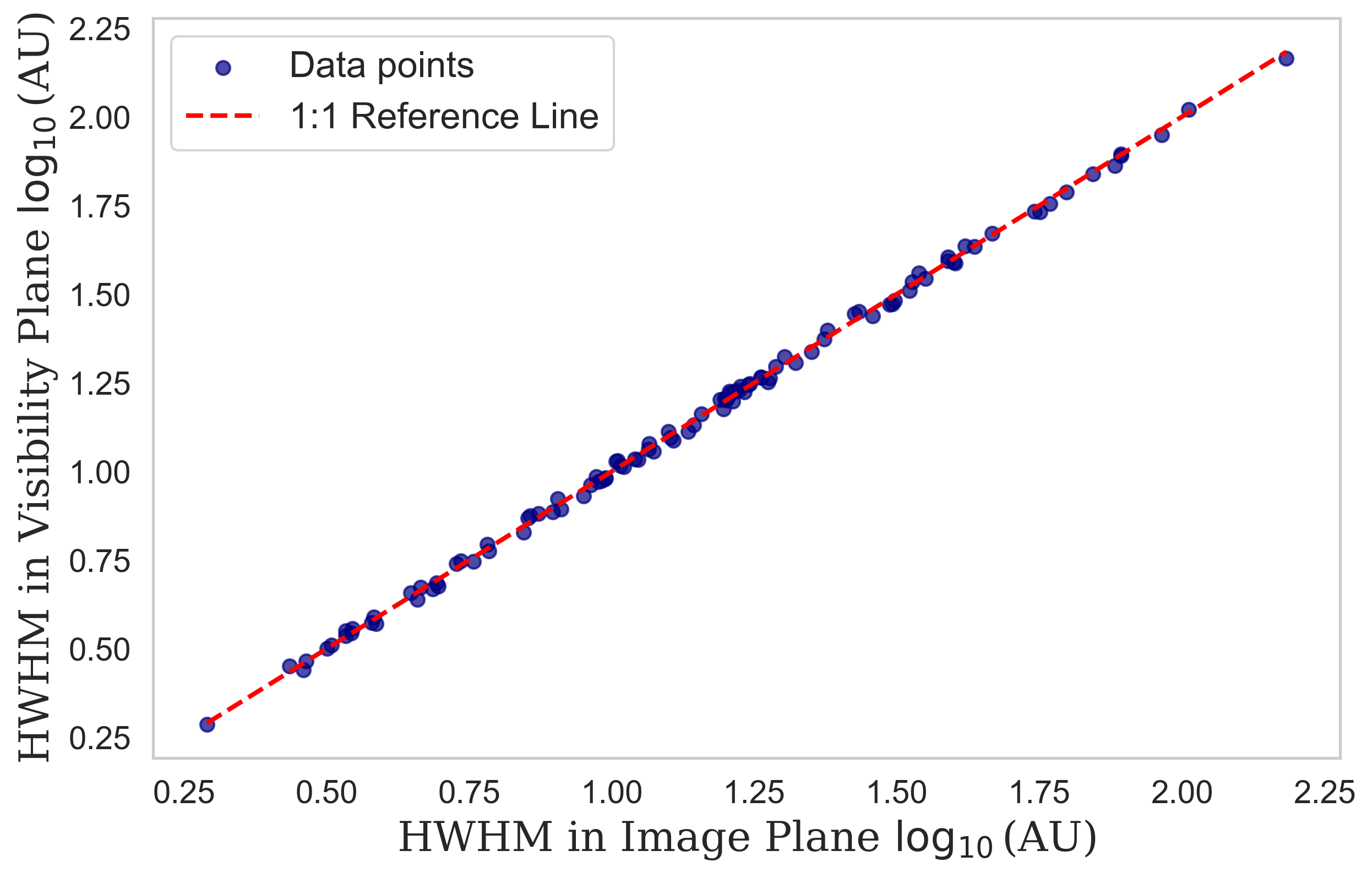}
        %\caption*{Top Left: A.png}
    \end{minipage}
    \hfill
    \begin{minipage}[b]{0.45\textwidth}
        \centering
        \includegraphics[width=\textwidth]{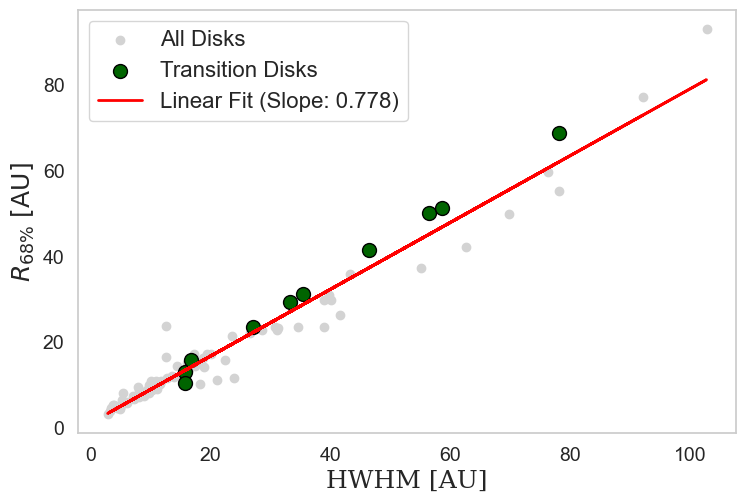}
        %\caption*{Top Left: A.png}
    \end{minipage}
    
    \caption{The left panel shows the disk radii fitted in the visibility plane (from  \textit{uvmodelfit)} vs. the disk radii measured in the image plane (from  \textit{imfit}). Both values agree within 3$\%$ across the full range. The right panel is the comparison between the  R$_{68\%}$ size estimates obtained from the Frank profiles and the HWHM values obtained from the Gaussian \textit{imfit} values.  The best fit lin (in red) has a slope of 0.78.  Objects with inner dust cavities (transition disks) are indicated and mostly fall above the line.}
    \label{App_fig}
\end{figure}

\begin{figure}[hb!]
    \centering
    \begin{minipage}[b]{0.8\textwidth}
        \centering
        \includegraphics[width=\textwidth]{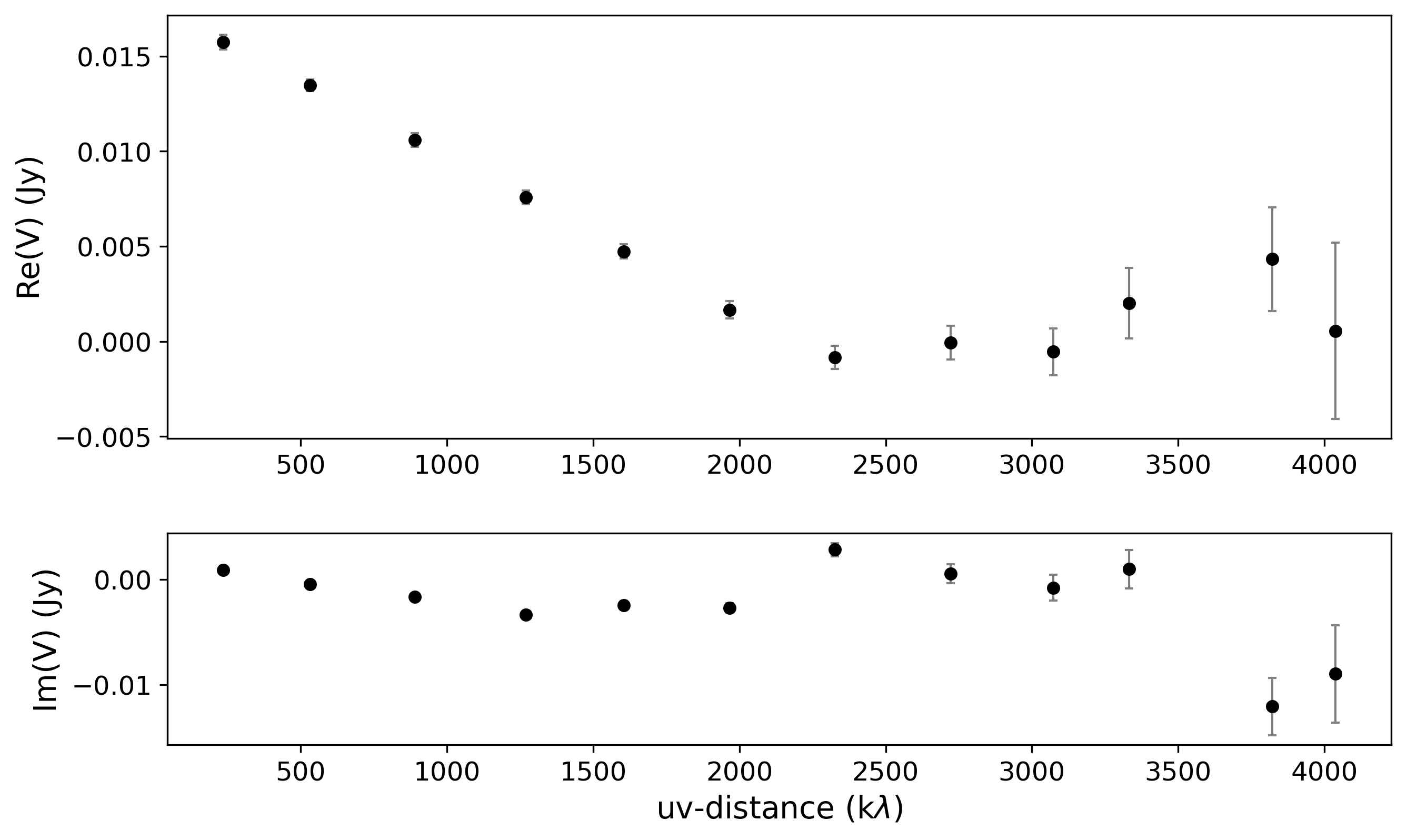}
        \caption{The deprojected visibility profile of one of the smallest objects (ODISEA\_C4\_099), with an estimated HWHM size of 21 mas and a 
        R$_{68\%}$ of 29 mas. The visibility profile falls with baseline and is therefore not consistent with a point source.  The sources is $\sim$ 16 mJy in Band 8 and   $\sim$ 5 mJy in Band 6.}
        \label{Sigma-Z_relation}
    \end{minipage}
\end{figure}

\begin{figure}[h]
    \centering
 \begin{minipage}[b]{0.45\textwidth}
        \centering
        \includegraphics[width=\textwidth]{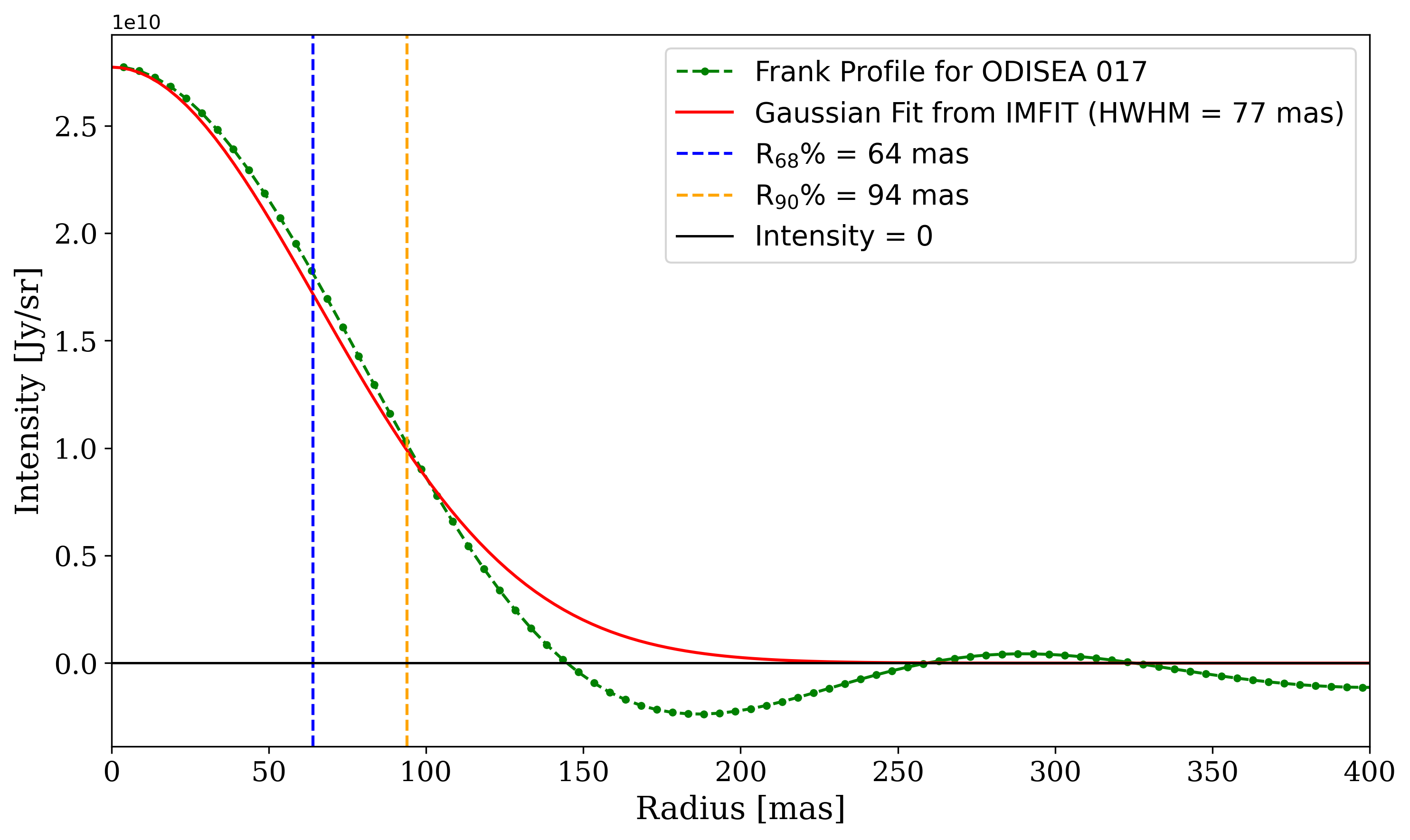}
        %\caption*{Top Left: A.png}
    \end{minipage}
    \hfill
    \begin{minipage}[b]{0.45\textwidth}
        \centering
        \includegraphics[width=\textwidth]{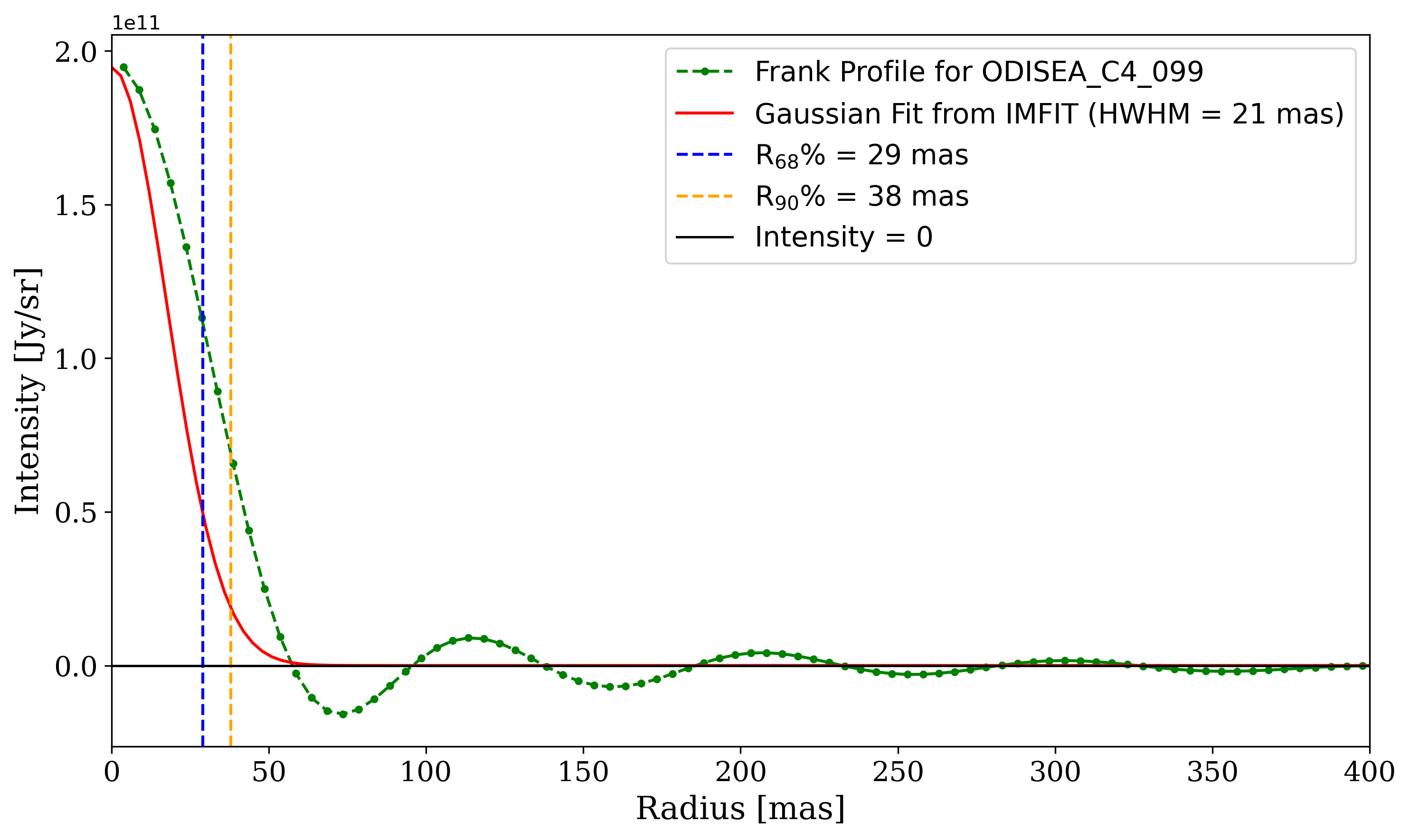}
        %\caption*{Top Left: A.png}
    \end{minipage}

    \caption{A comparison of the the profile obtained from Frank to the Gaussian fit from \textit{imfit} for two sources.  For well-resolved sources, \textit{imfit} reproduces well the intensity profile.  For very small sources, \textit{imfit}  might underestimate their size and Frank sizes should be preferred.}
    
    \label{App_fig-3}

\end{figure}

\begin{figure}[h]
    \centering
 \begin{minipage}[b]{0.45\textwidth}
        \centering
        \includegraphics[width=\textwidth]{Paper_Plots/cumulative_distribution_flux_with_error_shading.png}
        %\caption*{Top Left: A.png}
    \end{minipage}
    \hfill
    \begin{minipage}[b]{0.45\textwidth}
        \centering
        \includegraphics[width=\textwidth]{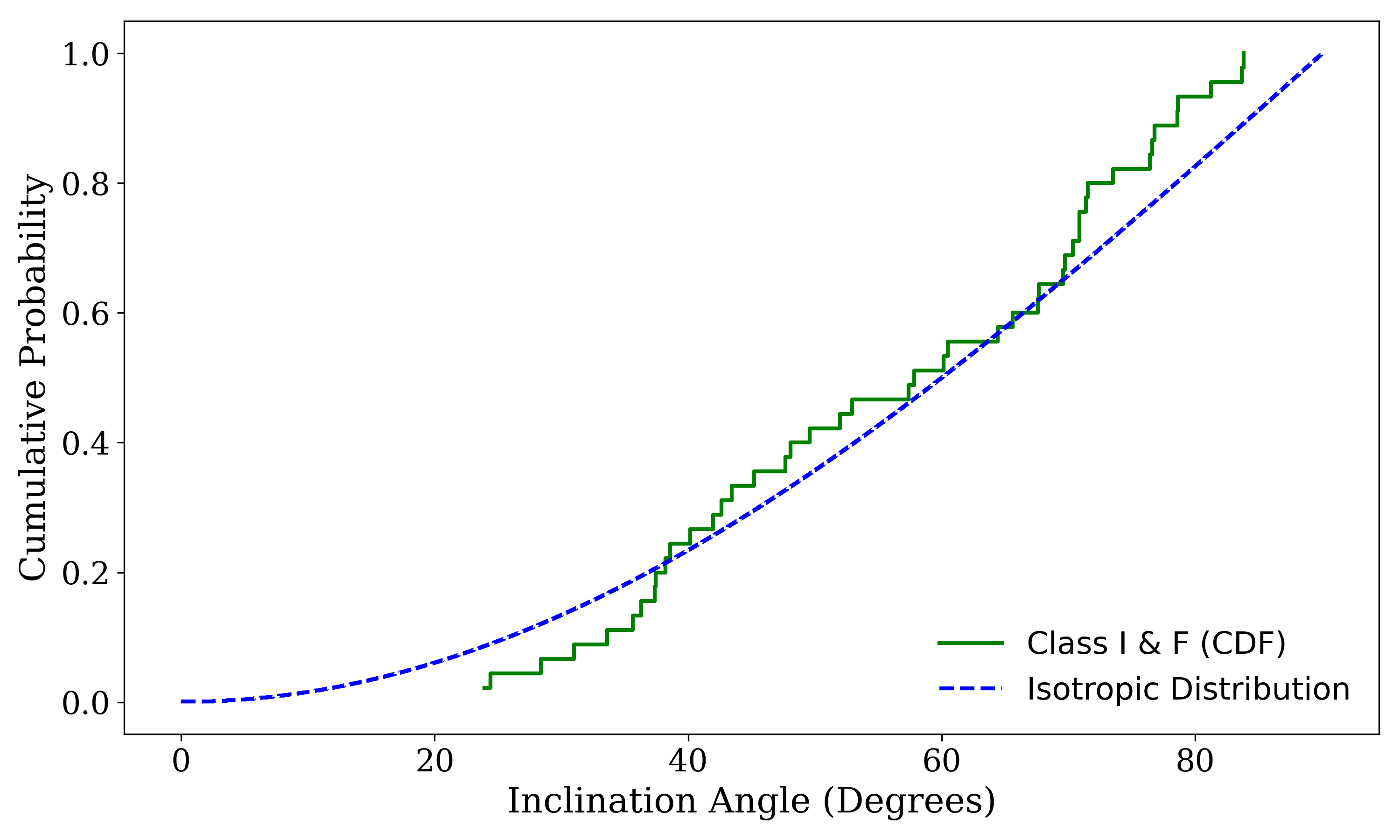}
        %\caption*{Top Left: A.png}
    \end{minipage}
    
    \caption{
    The left panel shows the cumulative flux distribution of Class II and embedded sources. The errors (shaded regions) correspond to the uncertainties in the Gaussian fits reported in Table 2. The flux distributions are indistinguishable from each other (see results of K-S test in Table 3). The right panel shows the cumulative distribution of disk inclinations for embedded sources (calculated from the ratios of semi-major to semi-minor axes) compared to the isotropic distribution (constructed from  1000 values with uniform distribution in inverse-cosine). The observed distribution of inclinations is also indistinguishable from a random distribution  (see K-S Test results in Table 3), suggesting that embedded sources are not significantly biased toward highly-inclined disks. The inclination are not expected to follow a Gaussian distribution and hence no Gaussian fit has been included.    
   }
    \label{App_fig-2}

\end{figure}
%\section{Appendix information}

\bibliography{References}{}
\bibliographystyle{aasjournal}

\end{document}